\documentclass[aps, prd, superscriptaddress, preprintnumbers]{revtex4}

\usepackage{graphicx,amsfonts,amsmath,amssymb,amstext}
\usepackage{float,wrapfig}
\usepackage{subfigure, psfrag}
\usepackage{dsfont}
\usepackage{color}
\usepackage{bm}
\usepackage{hyperref}
\hypersetup{
	colorlinks = true,
	linkcolor = blue,
	citecolor = blue
}

\newcommand{\be}{\begin{equation}}
\newcommand{\ee}{\end{equation}}
\newcommand{\ba}{\begin{eqnarray}}
\newcommand{\ea}{\end{eqnarray}}

\definecolor{purple}{rgb}{0.8,0,0.6}
\definecolor{darkgreen}{rgb}{0.00,0.6,0.00}

\begin{document}
\title{Fermi arcs and DC transport in nanowires of Dirac and Weyl semimetals}
\date{January 09, 2020}

\author{P.~O.~Sukhachov}
\email{pavlo.sukhachov@su.se}
\affiliation{Nordita, KTH Royal Institute of Technology and Stockholm University, Roslagstullsbacken 23, SE-106 91 Stockholm, Sweden}

\author{M.~V.~Rakov}
\affiliation{Institut f\"ur Mathematische Physik, Technische Universit\"at Braunschweig, Mendelssohnstra\ss e 3, 38106 Braunschweig, Germany}
\affiliation{Faculty of Physics, Kyiv National Taras Shevchenko University, 64/13 Volodymyrska st., 01601 Kyiv, Ukraine}

\author{O.~M.~Teslyk}
\affiliation{Faculty of Physics, Kyiv National Taras Shevchenko University, 64/13 Volodymyrska st., 01601 Kyiv, Ukraine}

\author{E.~V.~Gorbar}
\affiliation{Faculty of Physics, Kyiv National Taras Shevchenko University, 64/13 Volodymyrska st., 01601 Kyiv, Ukraine}
\affiliation{Bogolyubov Institute for Theoretical Physics, 03680 Kyiv, Ukraine}

\begin{abstract}
The transport properties and electron states in cylinder nanowires of Dirac and Weyl semimetals are studied paying special attention to the structure and properties of the surface Fermi arcs. The latter make the electric charge and current density distributions in nanowires strongly nonuniform as the majority of the charge density is accumulated at the surface. It is found that a Weyl semimetal wire also supports a magnetization current localized mainly at the surface because of the Fermi arcs contribution. By using the Kubo linear response approach, the direct current (DC) conductivity is calculated and it is found that its spatial profile is nontrivial. By explicitly separating the contributions of the surface and bulk states, it is shown that when the electric chemical potential and/or the radius of the wire is small, the electron transport is determined primarily by the Fermi arcs and the electrical conductivity is much higher at the surface than in the bulk. Due to the rise of the surface-bulk transition rate, the relative contribution of the surface states to the total conductivity gradually diminishes as the chemical potential increases. In addition, the DC conductivity at the surface demonstrates noticeable peaks when the Fermi level crosses energies of the surface states.
\end{abstract}

\maketitle

\section{Introduction}
\label{sec:Introduction}

Dirac and Weyl semimetals are condensed matter materials, whose low-energy electron quasiparticles are described by the relativistic-like Dirac and Weyl equations, respectively (for recent reviews on Dirac and Weyl semimetals, see Refs.~\cite{Yan-Felser:2017-Rev,Hasan-Huang:rev-2017,Armitage-Vishwanath:2017-Rev}). One of the hallmark properties of their quasiparticles is the linear energy dispersion relation in the vicinity of the band-touching points known as the Dirac points and Weyl nodes, respectively. The electron states in Weyl semimetals are topologically nontrivial, which is reflected in the fact that the Weyl nodes are monopoles of the Berry flux in momentum space~\cite{Berry:1984} with the topological charge directly related to the Weyl node chirality.

As was shown by Nielsen and Ninomiya~\cite{Nielsen-Ninomiya-1,Nielsen-Ninomiya-2}, simple continuity arguments imply that Weyl nodes can exist only in pairs of opposite chiralities. Since the lines of the Berry curvature cannot exit the topological semimetal or simply terminate at its surface, special topologically protected surface Fermi arc states connecting the projections of the bulk Weyl nodes onto the surface should necessarily exist. These states were theoretically predicted in Ref.~\cite{Savrasov:2011} and, as shown in Ref.~\cite{Haldane:2014}, provide a unique means to equilibrate the chemical potential in otherwise disconnected Weyl nodes of opposite chiralities. Experimentally, the Fermi arcs were directly observed with the help of the angle-resolved photoemission spectroscopy
~\cite{Xu-Hasan:2015,Xu-Hasan-TaP:2015,Lv-Ding-TaAs:2015,Lv-Ding-TaAs:2015b,Yang-Chen-TaAs:2015,Xu-Shi-TaP:2015,Xu-Hasan-NbAs:2015,Liu-Felser:2015,Xu-Hasan-TaAs:2016,Belopolski-Hasan:2016} as well as through the quasiparticle interference patterns~\cite{Gyenis-Bernevig:2016,Batabyal-Felser:2016,Inoue-Bernevig:2016,Zheng-Hasan:2016}. The experimental observation of the Fermi arcs is reviewed in Refs.~\cite{Hasan-Huang:rev-2017,Zheng-Hasan:rev-2018}.

The Fermi arcs appear when the surface is parallel to the chiral shift $\mathbf{b}$, which is the vector that separates the Weyl nodes of opposite chiralities in momentum space. In such a case, the Weyl nodes are projected onto the different points in the surface Brillouin zone. Although the Fermi arcs are topologically protected, their shape depends on the boundary conditions~\cite{Hosur:2012,Sun-Yan:2015,Xu-Sun-Co3Sn2S2:2018,Morali-Beidenkopf-Co3Sn2S2:2019,Yang-Chen-NbAs:2019}. Therefore, it is interesting what happens with the Fermi arc states if the surface is still parallel to the chiral shift but is, however, curved. A cylindric wire whose axis is directed along the chiral shift is one of the simplest geometries to study this question.

Since the Dirac point is composed of Weyl nodes of opposite chirality and is usually topologically trivial, one would not expect any Fermi arcs in 3D Dirac semimetals. However, as was demonstrated numerically~\cite{Wang:2012,Wang:2013}, the 3D Dirac semimetals $A_3$Bi ($A$=Na, K, Rb) and Cd$_3$As$_2$ possess nontrivial surface Fermi arcs. Later, these surface states were also found experimentally in Na$_3$Bi~\cite{Xu-Hasan-Na3Bi:2015}.
Two of us showed~\cite{Gorbar-Sukhachov:2015-Z2} that the underlying physical reason for the existence of the surface Fermi arcs in Dirac semimetals is connected with a discrete up-down parity symmetry of the low-energy effective Hamiltonian. As a result, all electron states are split into two separate sectors, each describing a Weyl semimetal with a pair of Weyl nodes and broken time-reversal ($\mathcal{T}$) symmetry. The time-reversal symmetry is preserved in the complete theory because it interchanges states from the two different sectors. Therefore, the corresponding Dirac semimetal can be identified as a $\mathbb{Z}_2$ Weyl semimetal.
The surface Fermi arc states in a semi-infinite slab of $\mathbb{Z}_2$ Weyl semimetals $A_3$Bi ($A$=Na, K, Rb) were studied in Ref.~\cite{Gorbar-Sukhachov:2015-Z2-FA} by employing a continuum low-energy effective model and also were investigated in slab geometries in Refs.~\cite{Potter:2014,Gorbar:2014qta,Molina:2017,Molina:2018}. In addition to the surface states, the nontrivial $\mathbb{Z}_2$ topology is manifested too in the motion of electron wavepackets in these materials~\cite{Gorbar:2018ynb}.

The transport properties of Weyl semimetal nanowires have been already studied by using the scattering matrix approach and numerical methods in Refs.~\cite{Baireuther-Beenakker:2016,Igarashi:2017,Kaladzhyan-Bardarson:2019}. Remarkably, it was shown that the contribution of the Fermi arc states to the electric current induced by a slowly varying magnetic field can be significant and even comparable to that of the bulk states regardless of the system size~\cite{Baireuther-Beenakker:2016}. In fact, the corresponding contribution resembles the current of the chiral magnetic effect~\cite{Fukushima:2008,Chang-Yang:2015,Burkov:rev-2015} and equals $-e^2/(4\pi^2\hbar^2c) \mu_5 B$, where $-e$ is the electron charge, $\mu_5$ is the chiral chemical potential, and $B$ is a magnetic field. Such a counterintuitive behavior is explained by the fact that the small number of surface states is compensated by their increased sensitivity to magnetic field. A similar conclusion about large contribution of the surface states was also phenomenologically reached in Ref.~\cite{Breitkreiz-Brouwer:2019} in the case of weak surface-bulk scattering.

Recently, it was explicitly demonstrated~\cite{Kaladzhyan-Bardarson:2019} that the existence of the surface and bulk states allows for two transport regimes. In the surface regime, the current is carried by the Fermi arc surface states with conductance increasing in steps as a function of the electric chemical potential $\mu$ when it lies in the bulk confinement gap due to the finite size of a sample. For highly doped samples, another bulk-surface transport regime takes place where the conductance $G$ is dominated by the bulk states and is quadratic in the electric chemical potential. The latter dependence is caused by the fact that the group velocity of quasiparticles is constant unlike the case of usual metals where $G \sim \mu^{3/2}$.
The interplay of magnetic field and surface states in the magnetotransport of Weyl semimetals was also investigated in Refs.~\cite{Ominato:2016,Wang:2017-QHE,Kaladzhyan-Bardarson:2019}.
Among other interesting effects, the presence of the Fermi arcs allows for the closed surface-bulk orbits when a magnetic field is perpendicular to a slab of Weyl semimetal~\cite{Potter:2014,Zhang-Vishwanath:2016}. These orbits were experimentally identified via the quantum oscillations measurements in Ref.~\cite{Moll:2016}. The transport evidence of the Fermi arc surface states in Dirac semimetal Cd$_3$As$_2$ nanowires was also demonstrated via the Aharonov--Bohm oscillations in Ref.~\cite{Wang-Liao:2016}. By using the concept of the surface-bulk orbits, the authors of Ref.~\cite{Baum:2015} investigated a nonlocal transport in a Weyl semimetal slab. In such a case, perturbation on, e.g., a top surface of Weyl semimetal produces a signal on the other surface.

Because of the topological nature of the Fermi arcs and their effectively 1D character with a linear energy dispersion, one would naively expect that the surface transport should be nondissipative. However, as was shown in Ref.~\cite{Gorbar-Sukhachov:2016}, this is generically not the case and, in fact, the Fermi arc transport is dissipative.
Indeed, since the gapless bulk states in Weyl semimetals coexist with the Fermi arc states, the latter are not fully decoupled from the bulk and there is a scattering into the bulk states in addition to possible scattering into other surface Fermi arc states.
This is in contrast to the case of topological insulators, in which bulk states are gapped and an effective Hamiltonian for the surface states can be reliably formulated.
In topological semimetals, the electron scattering from the surface into the bulk and vice versa leads to a dephasing of the Fermi arc quasiparticles and, consequently, dissipation. The dissolution of Fermi arcs in the presence of a strong disorder was also confirmed numerically in Refs.~\cite{Slager:2017,Wilson:2018}.

The main goal of this paper is to investigate the interplay of the finite size effects and the Fermi arc states in Dirac semimetals paying a special attention to the electric charge and current density profiles. The structure of the surface states, their energy dispersion, and wave functions are also rigorously investigated. Among the key results of this paper is the spatial profile of the DC conductivity in cylindrical nanowires, which, to the best of our knowledge, was not investigated before.
We believe that our study is important in a rapidly developing field of nanoelectronics involving Dirac and Weyl materials. For example, the use of Dirac semimetal Cd$_3$As$_2$ nanowires in creating controllable p-n junctions was recently demonstrated in Ref.~\cite{Bayogan-Jung:2019}.

The paper is organized as follows. The model of a Dirac or Weyl semimetal nanowire is defined in Sec.~\ref{sec:model}. The wave functions and the energy spectrum are discussed in the same section. Sec.~\ref{sec:Distribution-cc} is devoted to the electric charge and current density distributions inside nanowires. The DC response is investigated in Sec.~\ref{sec:Kubo}. The results are discussed and summarized in Sec.~\ref{sec:Summary}. Technical details related to the derivation of wave functions inside and outside the wire as well as the corresponding boundary conditions are presented in Appendices~\ref{sec:app-wave-functions} and \ref{sec:app-BC}, respectively. Conductivities for wires with different radiuses are presented in Appendix~\ref{sec:app-sigma-few-R}. A few useful formulas and relations are presented in Appendix~\ref{sec:app-Kubo-formulas}. In addition, we set the Planck and Boltzmann constants to unity, $\hbar=1$ and $k_{\rm B}=1$.

\section{Model, wave functions, and energy spectrum}
\label{sec:model}

In this section, the key details of Dirac and Weyl semimetal nanowires including the Hamiltonian, wave functions, and energy spectrum are discussed.

\subsection{Hamiltonian and boundary conditions}
\label{sec:model-H-BC}

Let us begin with the low-energy Hamiltonian of the electron states in the Dirac semimetals $A_3\mbox{Bi}$
($A$=Na, K, Rb)~\cite{Wang:2012} (which is also valid for one of the crystalline phases of Cd$_3$As$_2$~\cite{Wang:2013}). Its explicit form reads as
\begin{equation}
\label{model-H-Dirac}
H_{4\times4} = H_{2\times 2}^{+}\oplus H_{2\times 2}^{-},
\end{equation}
where the upper block is given by
\begin{equation}
\label{model-H-Dirac-up}
H_{2\times2}^{+}=(C_0+C_1k_z^2)\mathds{1}_2 +\left(
                                                  \begin{array}{cc}
                                                    \gamma\left(k_z^2-m\right) & v\left(k_x+ik_y\right) \\
                                                    v\left(k_x-ik_y\right) & -\gamma\left(k_z^2-m\right) \\
                                                  \end{array}
                                                \right)
\end{equation}
and the lower block $H_{2\times2}^{-}$ is obtained by replacing $k_x\to-k_x$ in $H_{2\times2}^{+}$. Note that, for simplicity, we
omitted terms $O(k_x^2)$ and $O(k_y^2)$. It is important to note that $m$ is positive in the Dirac and Weyl semimetals because, otherwise, there is no band-touching and a trivial insulator is realized (see the energy spectrum below and the discussion at the end of this
subsection).

Since the upper and lower blocks of Hamiltonian (\ref{model-H-Dirac}) do not mix in the model at hand, it is easy to find that each
of them has the following two-band energy spectrum:
\begin{eqnarray}
\label{model-H-Dirac-spectrum}
\epsilon_{\mathbf{k}} = (C_0+C_1k_z^2) \pm \sqrt{\gamma^2(m-k_z^2)^2 +v^2(k_x^2+k_y^2)}.
\end{eqnarray}
The bands touch at $\mathbf{k}_0^{\pm} =\left(0,0,\pm \sqrt{m}\right)$, which define the positions of the Dirac points (for the $4\times4$ model) or Weyl nodes (for each of the $2\times2$ blocks).
It is important to note that the upper block describes a $\cal{T}$ symmetry broken Weyl semimetal with the Weyl nodes of left- and right-handed chiralities located at $\mathbf{k}_0^{-}$ and $\mathbf{k}_0^{+}$, respectively. Since the lower block is obtained by replacing $k_x\to-k_x$, the chiralities of its Weyl nodes are reversed compared to those of the upper block. Therefore, the two copies of Weyl semimetals overlap and form a Dirac semimetal. The standard relativistic-like Hamiltonian can be straightforwardly obtained by expanding in the vicinity of the $\mathbf{k}_0^{\pm}$ points (see, e.g., Ref.~\cite{Gorbar-Sukhachov:2015-Z2-FA}).

\begin{figure*}[!ht]
\begin{center}
\includegraphics[width=0.45\textwidth]{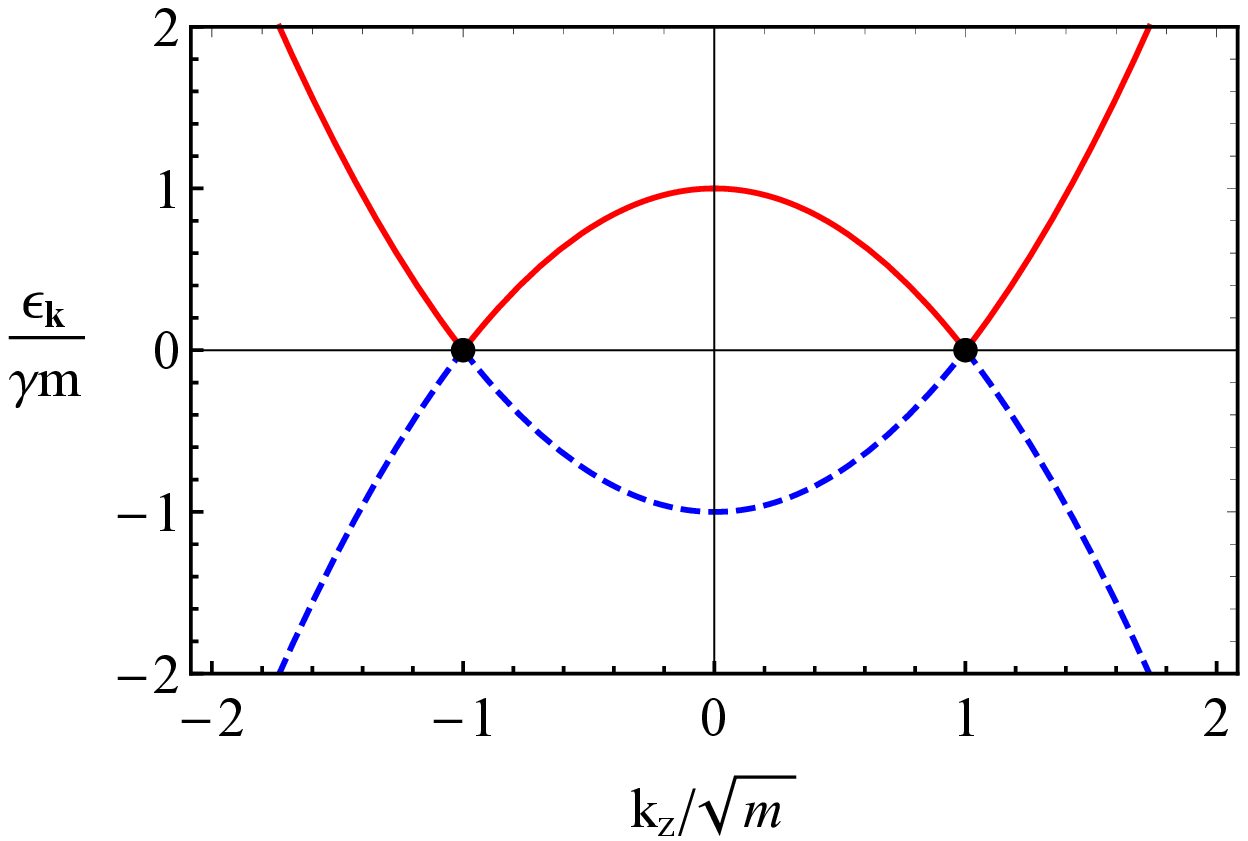}\hfill
\includegraphics[width=0.45\textwidth]{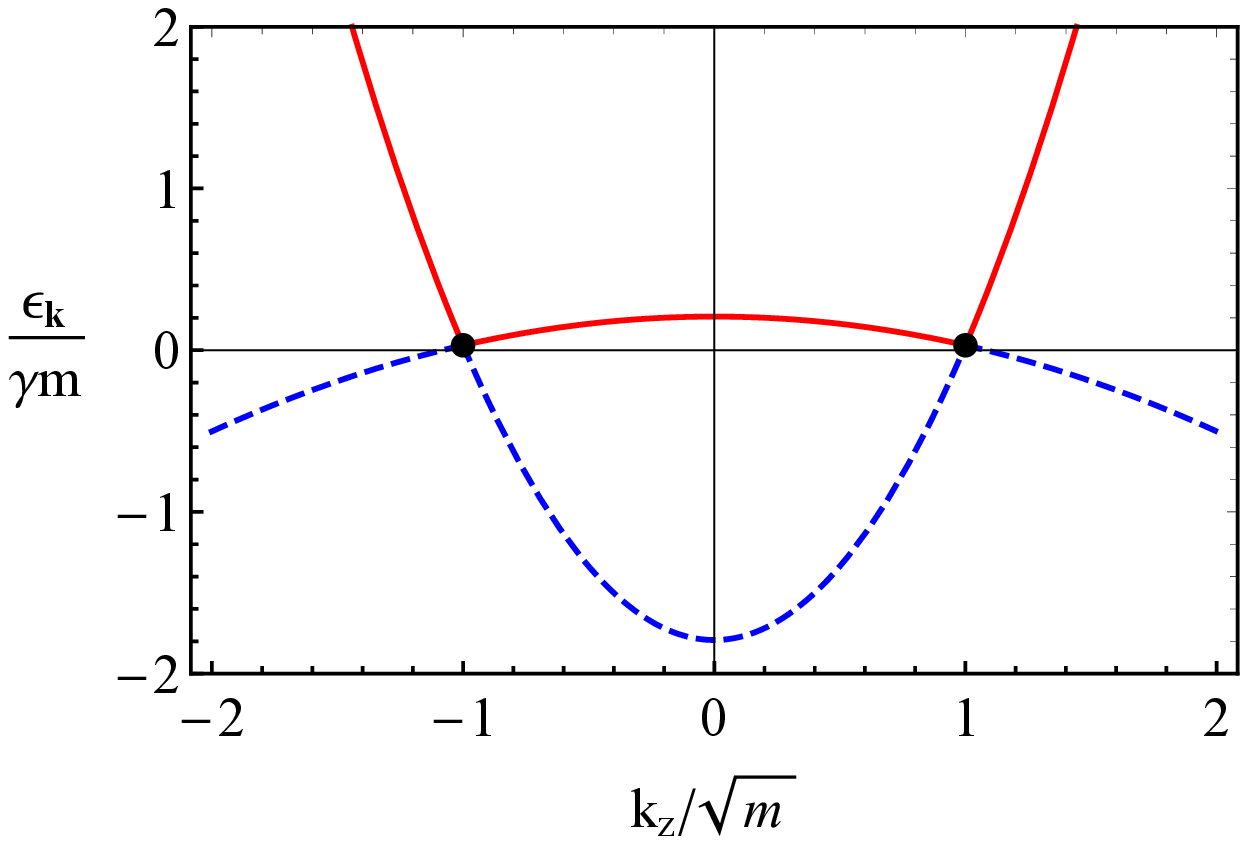}
\end{center}
\caption{The bulk energy spectrum (\ref{model-H-Dirac-spectrum}) of the $2\times 2$ model for $C_0=C_1=0$ (left panel) as well as $C_0$ and
$C_1$ defined in Eq.~(\ref{model-parameters}) (right panel).
}
\label{fig:model-spectrum-bulk}
\end{figure*}

By fitting the \emph{ab initio} numerical data for Na$_3$Bi with the effective Hamiltonian (\ref{model-H-Dirac}), the following values of model parameters were obtained in Ref.~\cite{Wang:2012}:
\begin{equation}
C_0 = -0.06382~\mbox{eV}, \quad C_1 = 8.7536~\mbox{eV\,\AA}^2, \quad m=0.008162~\mbox{\AA}^{-2}, \quad \gamma=10.6424~\mbox{eV\,\AA}^2, \quad  v=2.4598~\mbox{eV\,\AA}.
\label{model-parameters}
\end{equation}
We plot the bulk energy spectrum in the left and right panels of Fig.~\ref{fig:model-spectrum-bulk} for $C_0=C_1=0$ as well as
$C_0$ and $C_1$ defined in Eq.~(\ref{model-parameters}), respectively. It is clear that the term with $C_0$ and $C_1$ breaks the
particle-hole symmetry and plays the role of momentum-dependent electric chemical potential. In what follows, however, we will primarily
concentrate on the case $C_0=C_1=0$. We checked that, while these terms do affect the transport properties, they do not lead to any qualitatively new effects.

Since the main goal of our study is to consider the interplay of the finite-size effects and the Fermi arc surface states, we employ a simple model of a cylindrical wire of radius $R$, which allows for an analytical solution. The axis of the cylinder is parallel to the chiral shift, i.e., is directed along the $z$ axis. Furthermore, it is appropriate to use the cylindrical coordinates $\mathbf{r}=\left\{r\cos{\varphi}, r\sin{\varphi}, z\right\}$ in which the upper block of the Dirac Hamiltonian takes the following form in the mixed coordinate-momentum representation:
\begin{equation}
\label{model-cylinder-H-Dirac-up}
H_{2\times2}^{+} =\left(C_0+C_1k_z^2\right)\mathds{1}_2 + \left(
                   \begin{array}{cc}
                     \gamma(k_z^2-m) & -ive^{i\varphi}\left(\frac{\partial}{\partial r}+\frac{i}{r}\frac{\partial}{\partial \varphi}\right) \\
                     -ive^{-i\varphi}\left(\frac{\partial}{\partial r}-\frac{i}{r}\frac{\partial}{\partial \varphi}\right) & -\gamma(k_z^2-m) \\
                   \end{array}
                 \right).
\end{equation}
The lower block can be obtained by performing the complex conjugation in Hamiltonian~(\ref{model-cylinder-H-Dirac-up}), i.e., $H_{2\times2}^{-}=\left(H_{2\times2}^{+}\right)^{*}$.

The space outside the cylinder is vacuum or any trivial insulator that is modeled by the same Hamiltonian (\ref{model-H-Dirac}) albeit with $m\to -\tilde{m}$. As is easy to check by using the bulk spectrum (\ref{model-H-Dirac-spectrum}), this replacement opens the energy gap.
For simplicity, the energy gap $\tilde{m}$ outside the wire will be set to infinity.

\subsection{Wave functions and energy spectrum}
\label{sec:model-H-wf}

In this subsection, we determine the wave functions inside the wire as well as the energy spectrum. Since the operator of the
total angular momentum $\hat{J}_z= -i\partial_{\varphi} \mathds{1}_2 - \sigma_z/2$ commutes with Hamiltonian
(\ref{model-cylinder-H-Dirac-up}), we seek wave functions of Hamiltonian (\ref{model-cylinder-H-Dirac-up}) inside the wire ($r<R$) in the form
of eigenstates with angular momentum $n-1/2$, i.e.,
\begin{equation}
\label{model-cylinder-psi-Dirac-up-r<R}
\psi^{+}= \left(
                \begin{array}{c}
                  \rho_-(r) \, e^{in\varphi} \\
                  \rho_+(r) \, e^{i(n-1)\varphi} \\
                \end{array}
              \right).
\end{equation}
The wave functions for the lower block are $\psi^{-}=\left(\psi^{+}\right)^{*}$.

By solving the eigenvalue problem $H^{+}_{2\times 2}\psi^{+}=\epsilon\psi^{+}$, we find
(for the details of the derivation, see Appendix~\ref{sec:app-wave-functions})
\begin{equation}
\label{model-cylinder-psi-sp}
\psi_{+}^{+} =A_{+}(\tilde{\epsilon}) \left(
                      \begin{array}{c}
                        J_n\left(\frac{r}{r_0}\right) e^{in\varphi} \\
                        -i F(\tilde{\epsilon}) J_{n-1}\left(\frac{r}{r_0}\right) e^{i(n-1)\varphi}  \\
                      \end{array}
                    \right) 
\end{equation}
for $s_{\epsilon}>0$ and
\begin{equation}
\label{model-cylinder-psi-sm}
\psi_{-}^{+} =A_{-}(\tilde{\epsilon}) \left(
                      \begin{array}{c}
                        I_n\left(\frac{r}{r_0}\right) e^{in\varphi} \\
                        -i F(\tilde{\epsilon}) I_{n-1}\left(\frac{r}{r_0}\right) e^{i(n-1)\varphi}  \\
                      \end{array}
                    \right) 
\end{equation}
for $s_{\epsilon}<0$. Here $J_n(x)$ and $I_n(x)$ are the Bessel and modified Bessel functions of the first kind, respectively, and we used the
following shorthand notations:
\begin{eqnarray}
\label{model-cylinder-teps-def}
\tilde{\epsilon} &=& \epsilon-C_0-C_1k_z^2,\\
\label{model-cylinder-r0-def}
r_0 &=& \frac{v}{\sqrt{\left|\tilde{\epsilon}^2-\gamma^2(k_z^2-m)^2\right|}},\\
\label{model-cylinder-F-def}
F(\tilde{\epsilon}) &=& \frac{\sqrt{\left|\tilde{\epsilon}^{2}-\gamma^2(k_z^2-m)^2\right|}}{\tilde{\epsilon}+\gamma(k_z^2-m)},\\
\label{model-cylinder-spes-def}
s_{\epsilon} &=& \mbox{sgn}{\left[\tilde{\epsilon}^{2}-\gamma^2\left(k_z^2-m\right)^2\right]}.
\end{eqnarray}
The normalization constants are
\begin{eqnarray}
\label{model-cylinder-A-plus}
A_{+}(\tilde{\epsilon}) &=& \frac{1}{\sqrt{2\pi\left[\mathcal{J}(n,r_0)+F^{2}(\tilde{\epsilon})
\mathcal{J}(n-1,r_0)\right]}},\\
\label{model-cylinder-A-minus}
A_{-}(\tilde{\epsilon}) &=& \frac{1}{\sqrt{2\pi\left[\mathcal{I}(n,r_0) +F^{2}(\tilde{\epsilon})
\mathcal{I}(n-1,r_0)\right]}},
\end{eqnarray}
where
\begin{equation}
\label{model-cylinder-Jnr0-def}
\mathcal{J}(n,r_0)=\int_0^R r dr\,J_n^2\left(\frac{x}{r_0}\right) = \frac{R^2}{2}\left[J_n^2\left(\frac{R}{r_0}\right)+J_{n+1}^2\left(\frac{R}{r_0}\right)\right] -nRr_0J_n\left(\frac{R}{r_0}\right)J_{n+1}\left(\frac{R}{r_0}\right),
\end{equation}
\begin{equation}
\label{model-cylinder-Inr0-def}
\mathcal{I}(n,r_0)=\int_0^R r dr\,I_n^2\left(\frac{x}{r_0}\right) = \frac{R^2}{2}\left[I_n^2\left(\frac{R}{r_0}\right)-I_{n+1}^2\left(\frac{R}{r_0}\right)\right] -nRr_0I_n\left(\frac{R}{r_0}\right)I_{n+1}\left(\frac{R}{r_0}\right).
\end{equation}

In order to find energy levels, the wave function~(\ref{model-cylinder-psi-sp}) (or (\ref{model-cylinder-psi-sm}) for $s_{\epsilon}<0$) should be matched with the corresponding solution outside the cylinder.
As was discussed before, the latter is obtained by replacing $m\to -\tilde{m}$ and then setting $\tilde{m}\to\infty$. The
matching at the surface $r=R$ is considered in Appendix~\ref{sec:app-BC} and leads to the following boundary condition:
\begin{equation}
\label{model-cylinder-BC}
\rho_-(R)+i\rho_+(R)=0,
\end{equation}
which is valid for both upper and lower block solutions.

\begin{figure*}[!ht]
\begin{center}
\includegraphics[width=0.45\textwidth]{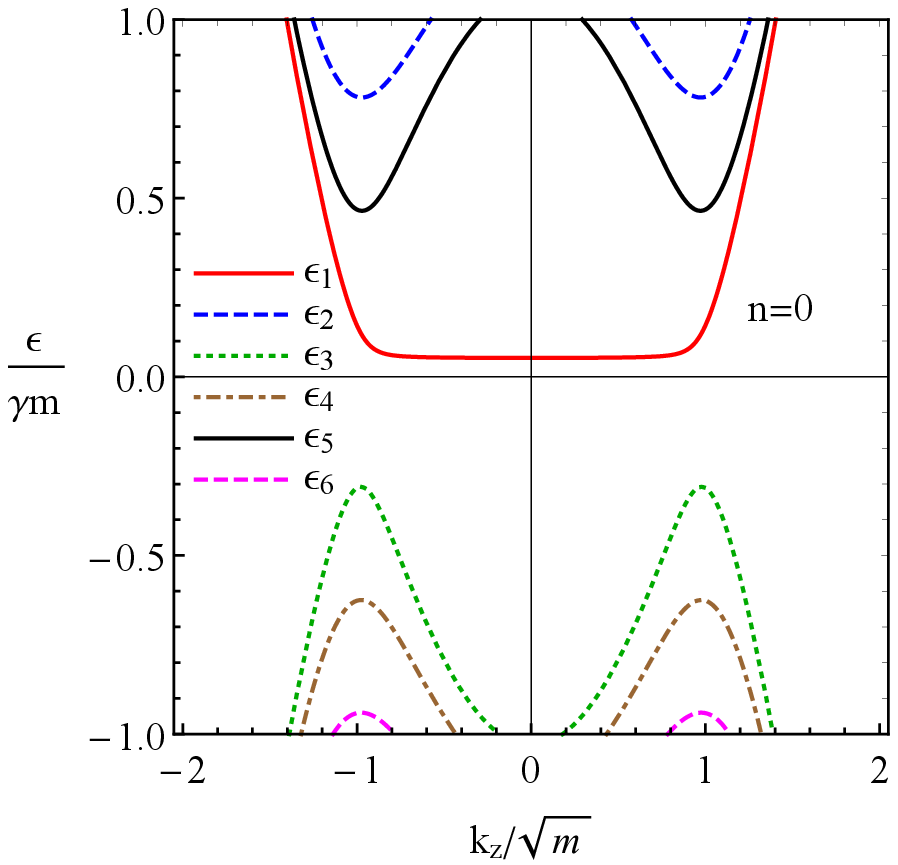}\hfill
\includegraphics[width=0.45\textwidth]{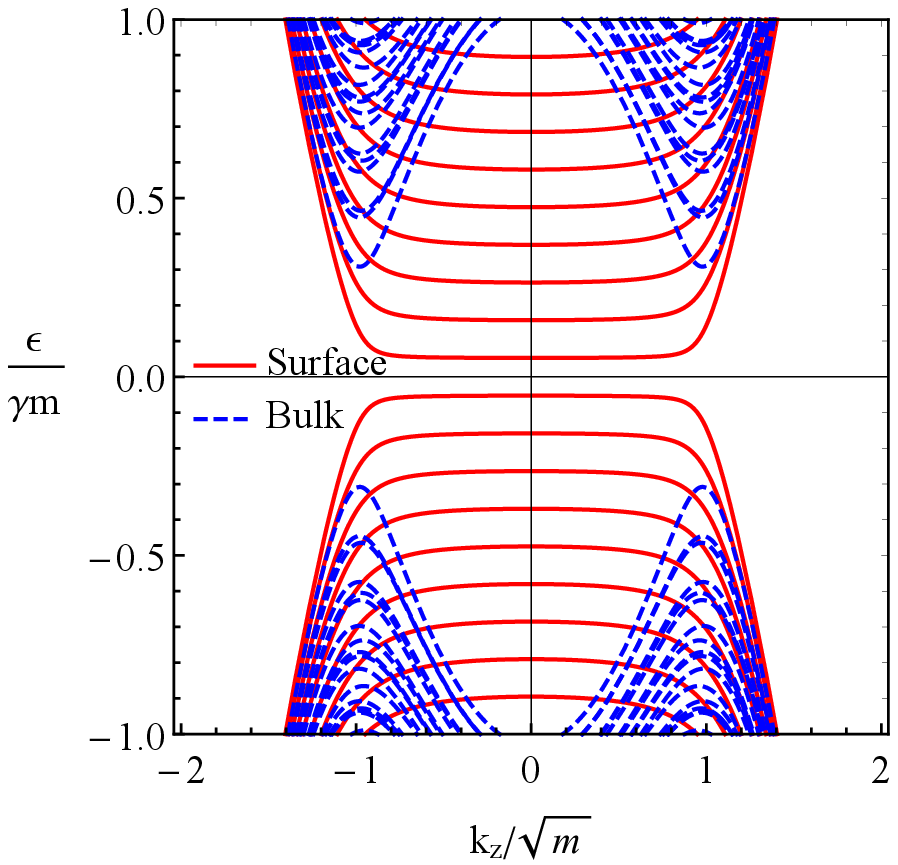}
\end{center}
\caption{The energy levels as a function of $k_z$ for the upper block at $R=10\,v/(\gamma m)$ and $C_0=C_1=0$ determined by
Eqs.~(\ref{spectrum-char-eq}) and (\ref{spectrum-char-eq-1}). While the left panel corresponds to $n=0$, the right one shows the results for $|n|<9$.
}
\label{fig:spectrum-2x2-large-R}
\end{figure*}

Let us find now the energy spectrum. By using the boundary condition (\ref{model-cylinder-BC}), we obtain the following characteristic
equation:
\begin{eqnarray}
\label{spectrum-char-eq}
s_{\epsilon}>0: &\quad& J_n\left(\frac{R}{r_0}\right) + F(\tilde{\epsilon})J_{n-1}\left(\frac{R}{r_0}\right)=0,\\
\label{spectrum-char-eq-1}
s_{\epsilon}<0: &\quad& I_n\left(\frac{R}{r_0}\right) + F(\tilde{\epsilon}) I_{n-1}\left(\frac{R}{r_0}\right)=0.
\end{eqnarray}
In general, these equations should be solved numerically.
We present the corresponding results for $C_0=C_1=0$ in the left and right panels of Fig.~\ref{fig:spectrum-2x2-large-R} for $n=0$ and $|n|<9$. If the electric chemical potential is fixed, then the sum over $n$ is naturally truncated because states with higher $n$ have larger energies.
Note also that in order to model realistic wires of Dirac and Weyl semimetals, we set $R=10\,v/(\gamma m)$, where $v/(\gamma m)\approx2.8319~\mbox{nm}$.
As was demonstrated in Refs.~\cite{Li-Yu-Cd3As2:2015,Wang-Liao:2016,Bayogan-Jung:2019}, the wires of such a radius can be synthesized, e.g., via the chemical vapour deposition method.

Our results reveal that there are two types of states: (i) modes with almost flat parabolic-like profiles, which correspond to the surface Fermi arc states (see the red solid lines in Fig.~\ref{fig:spectrum-2x2-large-R}), and (ii) modes with local extrema at $k_z =\pm \sqrt{m}$, which are the bulk states (see the blue dashed lines in Fig.~\ref{fig:spectrum-2x2-large-R}). With the growth of the wire radius the separation between the energy levels decreases until the Fermi arc levels merge and form a standard planar surface that connects separated bulk Fermi surfaces. The gap between the bulk states also reduces and vanishes at $R\to\infty$.
As was emphasized in Ref.~\cite{Kaladzhyan-Bardarson:2019}, while the Fermi arc and bulk states always coexist in large crystals, the finite-size effects make it possible to separate their contributions.
However, these effects themselves are not enough to probe the separate contributions of the surface and bulk modes. The other important ingredient is the position of the Fermi energy, which should be sufficiently low to intersect only the Fermi arc states but not the bulk states.

\section{Electric charge and current distributions}
\label{sec:Distribution-cc}

In this section, we consider the electric charge and current distributions inside nanowires of Dirac and Weyl semimetals.
In our study, we will pay a special attention to the separate contributions of the bulk and surface modes.

\subsection{Charge density}
\label{sec:Distribution-charge}

Let us start with the electric charge density. The principal quantity of interest is the probability density of each mode. Recall that since the wave functions of the lower and upper blocks are related by complex conjugation, the probability densities coincide for both blocks. In particular, they read as
\begin{eqnarray}
\label{Distribution-charge-psi-abs-plus-def}
\left|\psi^{\pm}_{+,n,k_z}(r)\right|^2 =  |A_{+}|^2\left[\left|J_n\left(\frac{r}{r_0}\right)\right|^2 +
\left|F(\tilde{\epsilon})\right|^2
\left|J_{n-1}\left(\frac{r}{r_0}\right)\right|^2 \right]
\end{eqnarray}
for $s_{\epsilon}>0$ and
\begin{eqnarray}
\label{Distribution-charge-psi-abs-minus-def}
\left|\psi^{\pm}_{-,n,k_z}(r)\right|^2 =  |A_{-}|^2\left[\left|I_n\left(\frac{r}{r_0}\right)\right|^2 +
\left|F(\tilde{\epsilon})\right|^2
\left|I_{n-1}\left(\frac{r}{r_0}\right)\right|^2 \right]
\end{eqnarray}
for $s_{\epsilon}<0$.

The electric charge density is defined through the sum of probability densities of eigenfunctions weighted with the equilibrium
electron distribution function $f^{\rm eq}(\epsilon)$
\begin{eqnarray}
\label{Distribution-charge-rho-def}
\rho^{\pm}(r)= -e\sum_{n=-\infty}^{\infty}\int\frac{dk_z}{2\pi} f^{\rm eq}(\epsilon)\left|\psi^{\pm}_{s_{\epsilon},n,k_z}(r)\right|^2,
\end{eqnarray}
where $-e$ is the charge of the electron and we use the standard Fermi--Dirac distribution function for the electron states ($\epsilon>0$) $f^{\rm eq}(\epsilon)=1/\left[1+e^{(\epsilon-\mu)/T}\right] \stackrel{T\to0}{=} \theta\left(\mu-\epsilon\right)$ and $f^{\rm eq}(\epsilon)=1-1/\left[1+e^{(\epsilon-\mu)/T}\right] \stackrel{T\to0}{=} 1-\theta\left(\mu-\epsilon\right)$ for the hole states ($\epsilon<0$). In the latter case one also needs to change $-e\to e$. In addition, here $T$ is temperature in the energy units.
At vanishing temperature, only the energy levels in the finite interval determined by the electric chemical potential $\mu$ contribute to the electric charge. It is clear that nonzero temperature will broaden this interval but should not lead to any new qualitative effects at least for sufficiently small temperature.

\begin{figure*}[!ht]
\begin{center}
\includegraphics[width=0.45\textwidth]{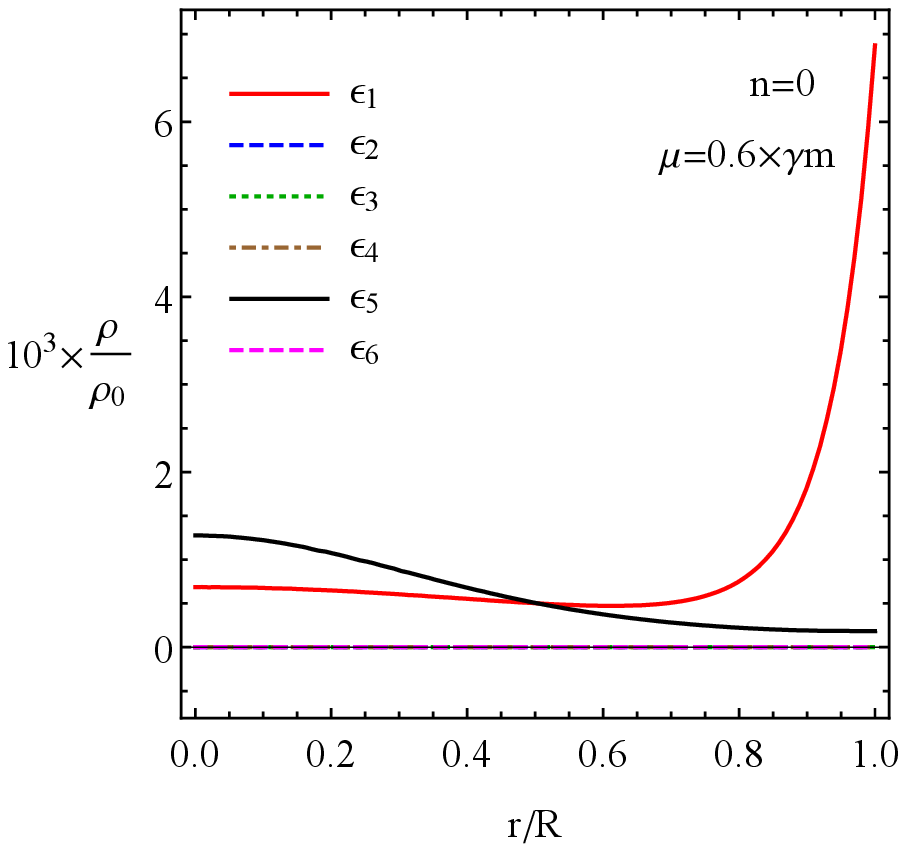}\hfill
\includegraphics[width=0.45\textwidth]{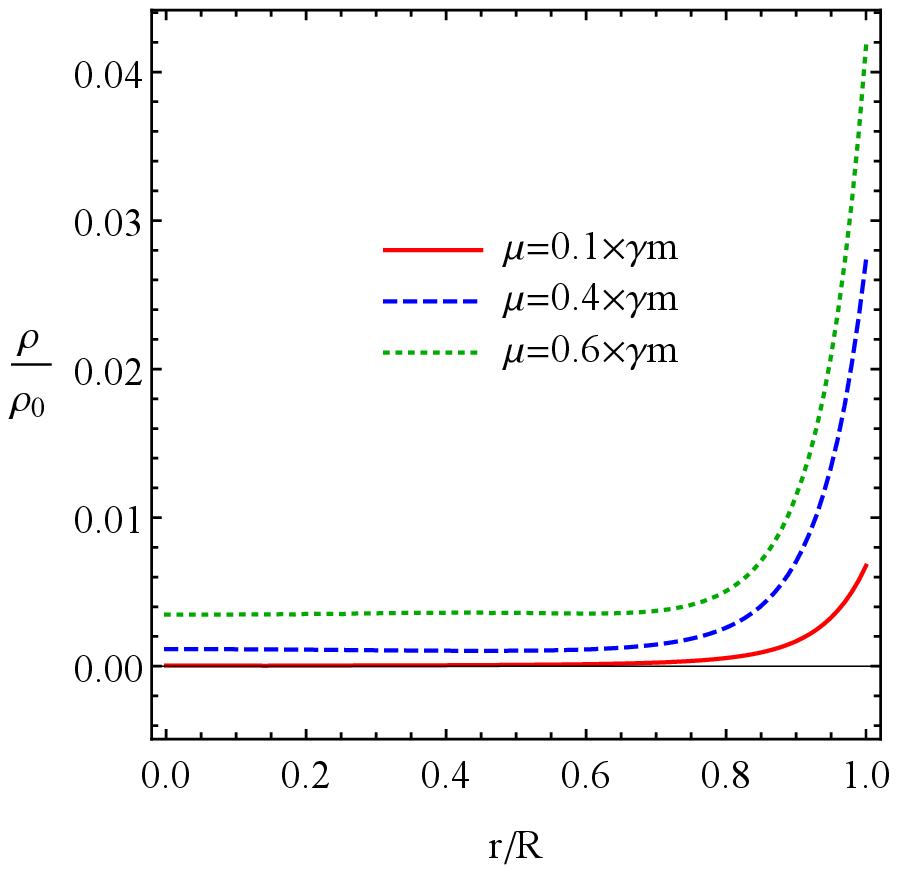}
\end{center}
\caption{The electric charge density given in Eq.~(\ref{Distribution-charge-rho-def}) inside a Weyl semimetal nanowire as a function of the radial coordinate $r$ at $n=0$ and $\mu=0.6\,\gamma m$ (left panel) and summed over $|n|<9$ at a few values of $\mu$ (right panel). The color code of the lines in the left panel is the same as in the left panel of Fig.~\ref{fig:spectrum-2x2-large-R}. In both panels, we set $R=10\,v/(\gamma m)$ and used $\rho_0 = -e\gamma^2 m^{5/2}/v^2$.}
\label{fig:2x2-charge-R=10}
\end{figure*}

The electric charge density defined by the $n=0$ mode for a Weyl semimetal nanowire of radius $R=10\,v/(\gamma m)$ is presented in the left panel of Fig.~\ref{fig:2x2-charge-R=10} at sufficiently large electric chemical potential $\mu=0.6\,\gamma m$. The color code of the lines is the same as in the left panel of Fig.~\ref{fig:spectrum-2x2-large-R}. As one can see, due to the Fermi arc states, the electric charge density is primarily accumulated at the surface (red solid line). The bulk states are also filled (black solid line) and the corresponding charge density is localized primarily near the center of the wire. However, their contribution is not dominant compared to the surface states.
The total charge density obtained by summing over $|n|<9$ is given in the right panel of Fig.~\ref{fig:2x2-charge-R=10}. As expected, for a small electric chemical potential, only the surface-localized Fermi arc modes contribute to the electric charge density. On the other hand, for large $\mu$, the contribution of the bulk states become noticeable, albeit still smaller than the surface one. Due to the equal contribution of both blocks, the charge density in a Dirac semimetal wire is doubled, i.e., $\rho(r)=\rho^{+}(r)+\rho^{-}(r)=2\rho^{+}(r)$. Note that this is not true when either time-reversal ${\cal T}$ or parity-inversion ${\cal P}$ symmetry is broken and an asymmetry between the $\mathbb{Z}_2$ copies of Weyl semimetals is present.

\subsection{Electric current distribution}
\label{sec:Distribution-current}

In this subsection, we consider the distribution of the electric current in nanowires of Dirac and Weyl semimetals.
In the absence of external fields, the current operator equals $\hat{\mathbf{j}} = -e\partial_{\mathbf{k}} H$. Then we find
\begin{eqnarray}
\label{Distribution-current-j-r-def}
\hat{j}_{r} &=& -ev\left[\cos{\varphi} \,\mathds{\tau}_z\otimes\sigma_x - \sin{\varphi} \,\mathds{1}_2\otimes\sigma_y\right],\\
\label{Distribution-current-j-phi-def}
\hat{j}_{\varphi} &=& -ev\left[-\sin{\varphi} \,\mathds{\tau}_z\otimes\sigma_x - \cos{\varphi} \,\mathds{1}_2\otimes\sigma_y\right],\\
\label{Distribution-current-j-z-def}
\hat{j}_z &=& -2ek_z\left[C_1 \,\mathds{1}_4 + \gamma \,\mathds{1}_2\otimes\sigma_z\right].
\end{eqnarray}
Here $\hat{j}_{r}$ and $\hat{j}_{\varphi}$ are the radial and azimuthal components of the electric current operator, respectively and $\hat{j}_z$ is the longitudinal current operator in the direction of the cylinder axis.

By using definitions (\ref{Distribution-current-j-r-def}) and (\ref{Distribution-current-j-phi-def}), we find
the following expectation values at $s_{\epsilon}>0$:
\begin{eqnarray}
\label{Distribution-current-j-r-calc}
\left[\psi_{+,n,k_z}^{\pm}(r)\right]^{\dag} \hat{j}_{r} \psi_{+,n,k_z}^{\pm}(r) &=& 0,\\
\label{Distribution-current-j-phi-calc}
\left[\psi_{+,n,k_z}^{\pm}(r)\right]^{\dag} \hat{j}_{\varphi} \psi_{+,n,k_z}^{\pm}(r) &=&
\mp 2ev|A_{+}|^2 F(\tilde{\epsilon}) J_{n-1}\left(\frac{r}{r_0}\right) J_{n}\left(\frac{r}{r_0}\right),\\
\label{Distribution-current-j-z-calc}
\left[\psi_{+,n,k_z}^{\pm}(r)\right]^{\dag} \hat{j}_{z} \psi_{+,n,k_z}^{\pm}(r) &=& -2ek_z|A_{+}|^2 \left[\left(C_1+\gamma\right) \left|J_n\left(\frac{r}{r_0}\right)\right|^2 +\left(C_1-\gamma\right) \left|F(\tilde{\epsilon})\right|^2 \left|J_{n-1}\left(\frac{r}{r_0}\right)\right|^2\right].
\end{eqnarray}
For $s_{\epsilon}<0$, one needs to replace $\psi_{+,n,k_z}\to \psi_{-,n,k_z}$, $A_{+}\to A_{-}$, and $J_n\left(r/r_0\right)\to I_n\left(r/r_0\right)$ in Eqs.~(\ref{Distribution-current-j-r-calc}), (\ref{Distribution-current-j-z-calc}),  and (\ref{Distribution-current-j-phi-calc}).
Explicit expressions for the coefficients $|A_{\pm}|^2$ are given in Eqs.~(\ref{model-cylinder-A-plus}) and (\ref{model-cylinder-A-minus}).

As expected, the radial component of the electric current expectation value vanishes everywhere in the wire.
As to the expectation value of the azimuthal component of the current, it has opposite signs for the upper and lower blocks. This means that
the total electric current density in the equilibrium state is zero as it should be in Dirac semimetals where the ${\cal T}$
symmetry is preserved.

\begin{figure*}[!ht]
\begin{center}
\includegraphics[width=0.45\textwidth]{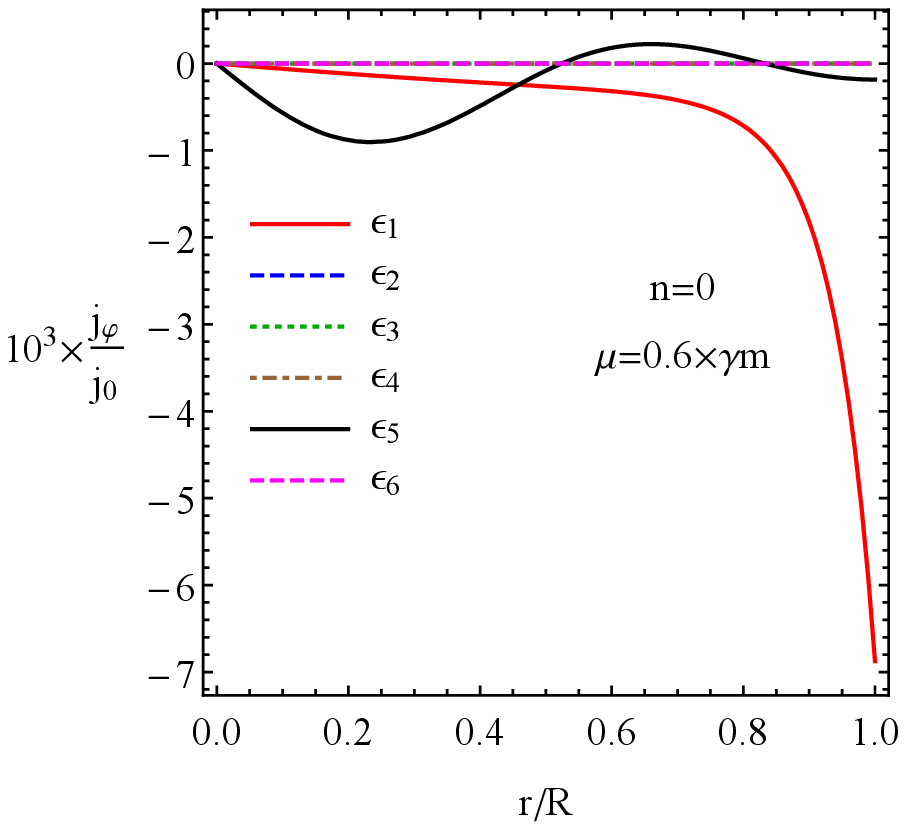}\hfill
\includegraphics[width=0.45\textwidth]{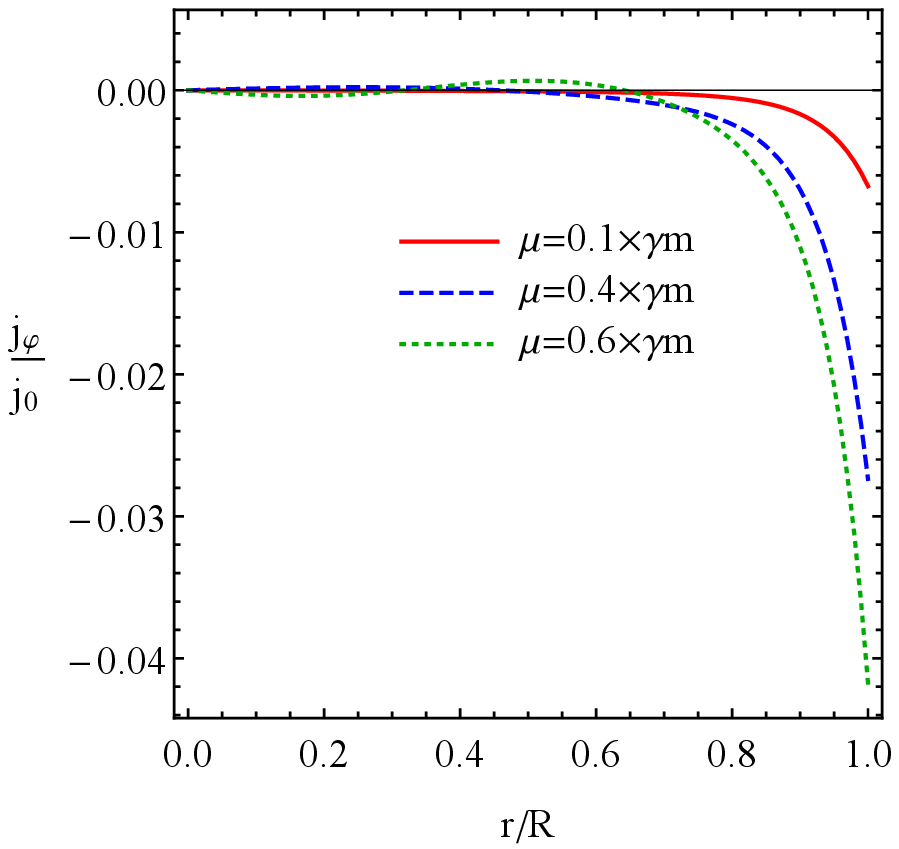}
\end{center}
\caption{
The azimuthal component of the electric current density given in Eq.~(\ref{Distribution-current-j-def}) inside the wire as a function of $r$ for $n=0$ and $\mu=0.6\,\gamma m$ (left panel) and summed over $|n|\leq9$ at a few values of $\mu$ (right panel). Only the upper block in Hamiltonian (\ref{model-H-Dirac}) is considered. The color code of the lines in the left panel is the same as in the left panel of Fig.~\ref{fig:spectrum-2x2-large-R}. In both panels, we set $R=10\,v/(\gamma m)$ and used $j_0 = -e\gamma^2 m^{3/2}/v$.}
\label{fig:2x2-j-phi-integrated-R=10}
\end{figure*}

Similarly to the electric charge density (\ref{Distribution-charge-rho-def}), the electric current density is defined as
\begin{eqnarray}
\label{Distribution-current-j-def}
\mathbf{j}^{\pm}(r)= -e\sum_{n=-\infty}^{\infty}\int\frac{dk_z}{2\pi} f^{\rm eq}(\epsilon)\left[\psi_{n,k_z}^{\pm}(r)\right]^{\dag} \hat{\mathbf{j}} \psi_{n,k_z}^{\pm}(r).
\end{eqnarray}
Since the energy spectrum is symmetric with respect to the replacement $k_z\to-k_z$, the longitudinal component of the electrical current density $j_z(r)$ vanishes after integrating over momentum.
The only nontrivial component of the electric current density is the azimuthal one, i.e., $j_{\varphi}(r)$.
Its dependence on $r$ for the upper block (i.e., for a $\mathcal{T}$ symmetry broken Weyl semimetal) and $R=10\,v/(\gamma m)$ is shown in Fig.~\ref{fig:2x2-j-phi-integrated-R=10}.
As one can see from the left panel, the main contribution to $j_{\varphi}(r)$ comes from the Fermi arc states (red solid line). The bulk states also contribute when $\mu$ is sufficiently large.
The corresponding contribution oscillates inside the wire and is smaller than that of the Fermi arcs.
The total current obtained after the summation over $|n|\leq9$ is shown in the right panel of Fig.~\ref{fig:2x2-j-phi-integrated-R=10}. As one can see, the surface contribution dominates even at sufficiently high electric chemical potential $\mu=0.6\,\gamma m$.
As we mentioned above, $j_{\varphi}(r)$ for the lower block in Hamiltonian (\ref{model-H-Dirac}) has the opposite sign compared to the azimuthal component of the electric current for the upper block. Therefore, as expected from the symmetry arguments, there is no magnetization current in Dirac semimetal nanowires.

\section{DC conductivity}
\label{sec:Kubo}

In this section the DC conductivity in nanowires of Dirac and Weyl semimetals is investigated in the Kubo linear response approach. While the DC response in nanowires was already investigated in Ref.~\cite{Baireuther-Beenakker:2016,Igarashi:2017,Kaladzhyan-Bardarson:2019}, the spatial profiles of the conductivity were not rigorously analyzed to the best of our knowledge.

\subsection{Kubo linear response approach}
\label{sec:Kubo-general}

The Kubo linear response theory is a standard method to determine the transport properties of various materials (see, e.g., Ref.~\cite{Bruus:2004-book,Mahan:book-2013}). In this approach, the DC conductivity tensor is defined through the imaginary part of the retarded current-current correlation function in the limit $\Omega \to 0$ (recall that the imaginary part of the conductivity is zero at $\Omega\to0$)
\begin{equation}
\sigma_{ij}(r, \varphi) = 
\lim_{\Omega \to 0}
\frac{\mbox{Im}\, \Pi_{ij}(\Omega+i0, q_z=0; r, \varphi)}{\Omega}.
\label{Kubo-sigma}
\end{equation}
The paramagnetic part of $\Pi_{ij}$ is given by
\begin{equation}
\Pi_{ij}(\Omega, q_z; r, \varphi) =  T \sum_{p=-\infty}^{\infty}
\int \frac{dk_z}{2\pi} \int_0^{2\pi} d\varphi^{\prime} \int_0^{R} r^{\prime}dr^{\prime} \mbox{tr} \left[ \hat{j}_{i}
G(i\omega_{p}, k_z; r, r^{\prime}, \varphi, \varphi^{\prime}) \hat{j}_{j}
G(i\omega_{p}-\Omega, k_z-q_z; r^{\prime}, r, \varphi^{\prime}, \varphi)\right],
\label{Kubo-Pi-ij-def}
\end{equation}
where $\omega_{p} =(2p+1)\pi T$ is the fermionic Matsubara frequency, $p$ is an integer, and the components of the current operator are defined in Eqs.~(\ref{Distribution-current-j-r-def}), (\ref{Distribution-current-j-phi-def}), and (\ref{Distribution-current-j-z-def}).
Note that the diamagnetic part of the retarded current-current correlation function, which originates from the quadratic dispersion along the $z$ axis, is irrelevant for the DC conductivity because it contributes to the imaginary part of $\sigma_{ij}$.

The retarded Green's function in the mixed coordinate-momentum representation is defined as
\begin{eqnarray}
\label{Kubo-G-def}
G(\omega, k_z; r, r^{\prime}, \varphi, \varphi^{\prime}) = i \sum_{n=-\infty}^{\infty} \sum_{l} \frac{\psi_{n,l}(r, \varphi) \psi_{n,l}^{\dag}(r^{\prime}, \varphi^{\prime}) }{\omega +\mu -\epsilon_{n,l}+i0},
\end{eqnarray}
where $n$ runs over all orbital angular momenta, $l$ denotes energy branches, and, for simplicity, we do not show the explicit dependence of $\epsilon_{n,l}$ on $k_z$, i.e., $\epsilon_{n,l}\equiv \epsilon_{n,l}(k_z)$. The explicit form of the retarded Green's function for the upper block and $s_{\epsilon}>0$ reads as
\begin{eqnarray}
\label{Kubo-G-exp}
G(\omega, k_z; r, r^{\prime}, \varphi, \varphi^{\prime}) = i \sum_{n=-\infty}^{\infty} \sum_{l} \frac{|A_{+}(\tilde{\epsilon}_{n,l})|^2}{\omega +\mu -\epsilon_{n,l}+i0}\, \hat{M}.
\end{eqnarray}
Here
\begin{eqnarray}
\label{Kubo-A-matrix}
\hat{M} = \left(
  \begin{array}{cc}
    J_n\left(\frac{r}{r_0}\right) J_n\left(\frac{r^{\prime}}{r_0}\right) e^{in(\varphi-\varphi^{\prime})} & i J_n\left(\frac{r}{r_0}\right) J_{n-1}\left(\frac{r^{\prime}}{r_0}\right) F\left(\tilde{\epsilon}_{n,l}\right) e^{in(\varphi-\varphi^{\prime}) +i\varphi^{\prime}} \\
    -i J_{n-1}\left(\frac{r}{r_0}\right) J_{n}\left(\frac{r^{\prime}}{r_0}\right) F\left(\tilde{\epsilon}_{n,l}\right) e^{-in(\varphi-\varphi^{\prime}) -i\varphi} & J_{n-1}\left(\frac{r}{r_0}\right) J_{n-1}\left(\frac{r^{\prime}}{r_0}\right) \left|F\left(\tilde{\epsilon}_{n,l}\right)\right|^2 e^{-i(n-1)(\varphi-\varphi^{\prime})} \\
  \end{array}
\right).
\end{eqnarray}
Furthermore, $F\left(\tilde{\epsilon}_{n,l}\right)$ and $|A_{+}|^2$ are given in Eqs.~(\ref{model-cylinder-F-def}) and (\ref{model-cylinder-A-plus}), respectively.
For $s_{\epsilon}<0$, one needs to replace the Bessel function $J_n(x)$ with the modified Bessel function $I_n(x)$ and $|A_{+}|^2\to |A_{-}|^2$ in Eq.~(\ref{Kubo-G-exp}).

Defining the spectral function $A(\omega, k_z, r, r^{\prime}, \varphi, \varphi^{\prime})$ as
\begin{eqnarray}
\label{Kubo-A-def}
A(\omega, k_z; r, r^{\prime}, \varphi, \varphi^{\prime}) = \frac{1}{2\pi} \left[ G_{\mu=0}(\omega+i 0, k_z; r, r^{\prime}, \varphi, \varphi^{\prime}) -G_{\mu=0}(\omega-i 0, k_z; r, r^{\prime}, \varphi, \varphi^{\prime}) \right],
\end{eqnarray}
one can re-express Green's function in the following equivalent form:
\begin{equation}
\label{Kubo-A-G}
G(\Omega, k_z; r, r^{\prime}, \varphi, \varphi^{\prime})  = i\int_{-\infty}^{\infty} d\omega \frac{A (\omega, k_z; r, r^{\prime}, \varphi, \varphi^{\prime})}{\Omega+\mu-\omega}.
\end{equation}
The explicit expression of the spectral function for the upper block and $s_{\epsilon}>0$ is readily obtained from Eq.~(\ref{Kubo-G-exp}) by replacing $i/\left[\omega +\mu -\epsilon_{n,l}+i0\right]$ with $\delta\left(\omega -\epsilon_{n,l}\right)$.

By using the spectral function representation (\ref{Kubo-A-def}), it is straightforward to obtain
\begin{eqnarray}
\label{Kubo-Pi-ij}
\Pi_{ij}(\Omega, q_z; r, r^{\prime}, \varphi, \varphi^{\prime}) &=& - \int\int  d\omega  d \omega^{\prime} \frac{f^{\rm eq}(\omega)-f^{\rm eq}(\omega^{\prime})}{\omega -\omega^{\prime}-\Omega-i0} \nonumber\\
&\times&\int \frac{dk_z}{2\pi} \int_0^{2\pi} d\varphi^{\prime} \int_0^{R} r^{\prime}dr^{\prime} \mbox{tr} \left[ \hat{j}_{i} A(\omega, k_z; r, r^{\prime}, \varphi, \varphi^{\prime}) \hat{j}_{j} A(\omega^{\prime}, k_z-q_z; r^{\prime}, r, \varphi^{\prime}, \varphi)\right].
\end{eqnarray}

The calculation of the longitudinal conductivity is rather simple because the trace in Eq.~(\ref{Kubo-Pi-ij}) is real. Then, by using the identity
\begin{equation}
\frac{1}{\omega -\omega^\prime-\Omega \mp i 0} = \mbox{p.v.}\frac{1}{\omega -\omega^\prime-\Omega} \pm i \pi \delta\left(\omega -\omega^\prime-\Omega\right),
\label{Kubo-P-value}
\end{equation}
where p.v. stands for the principal value,
as well as Eqs.~(\ref{Kubo-sigma}) and (\ref{Kubo-Pi-ij}), we straightforwardly derive the following longitudinal DC conductivity
\begin{eqnarray}
\label{Kubo-sigma-zz-DC}
\sigma_{zz}(r, \varphi) &=& \int d\omega \frac{4\pi e^2}{4T\cosh^2{\left(\frac{\omega-\mu}{2T}\right)}}
\int \frac{dk_z}{2\pi} \int_0^{2\pi} d\varphi^{\prime}  \int_0^{R} r^{\prime}dr^{\prime} \sum_{n} \sum_{l,l^{\prime}} k_z^2 \left|A_{s_{\epsilon}}(\tilde{\epsilon}_{n,l})\right|^2 \left|A_{s_{\epsilon^{\prime}}}(\tilde{\epsilon}_{n,l^{\prime}})\right|^2 \delta_{\Gamma}\left(\omega - \epsilon_{n,l}\right) \nonumber\\
&\times&  \delta_{\Gamma}\left(\omega- \epsilon_{n,l^{\prime}}\right) \left\{M_{11}M_{11}^{\prime} \left(\gamma+C_1\right)^2 +M_{22}M_{22}^{\prime}\left(\gamma-C_1\right)^2 +\left(C_1^2-\gamma^2\right) \left[M_{12}M_{21}^{\prime} +M_{21}M_{12}^{\prime}\right] \right\} \nonumber\\
&\stackrel{T\to0}{=}& 4\pi e^2 \int \frac{dk_z}{2\pi} \int_0^{2\pi} d\varphi^{\prime}  \int_0^{R} r^{\prime}dr^{\prime} \sum_{n} \sum_{l,l^{\prime}} k_z^2 \left|A_{s_{\epsilon}}(\tilde{\epsilon}_{n,l})\right|^2 \left|A_{s_{\epsilon^{\prime}}}(\tilde{\epsilon}_{n,l^{\prime}})\right|^2 \delta_{\Gamma}\left(\mu - \epsilon_{n,l}\right) \delta_{\Gamma}\left(\mu- \epsilon_{n,l^{\prime}}\right) \nonumber\\
&\times&  \left\{M_{11}M_{11}^{\prime} \left(\gamma+C_1\right)^2 +M_{22}M_{22}^{\prime}\left(\gamma-C_1\right)^2 +\left(C_1^2-\gamma^2\right) \left[M_{12}M_{21}^{\prime} +M_{21}M_{12}^{\prime}\right] \right\},
\end{eqnarray}
where we expanded to linear order in $\Omega$ and took the limit $\Omega\to0$. Here $M_{ij}$ denotes the $ij$ component of the spectral function matrix given in Eq.~(\ref{Kubo-A-matrix}) for the upper block and $s_{\epsilon}>0$. In the case of $M_{ij}^{\prime}$, one should replace $r_0\to r_0^{\prime}$ and $l\to l^{\prime}$, as well as $r\leftrightarrow r^{\prime}$ and $\varphi\leftrightarrow \varphi^{\prime}$. For the lower block, the spectral function matrix is $\hat{M}^*$.

It is important to include the effects of disorder in the study of the real part of the longitudinal conductivity. This can be done
phenomenologically through the replacement
\begin{equation}
\label{Kubo-delta-Gamma}
\delta\left(\mu - \epsilon_{n,l}\right) \to \delta_{\Gamma}\left(\mu - \epsilon_{n,l}\right) =  \frac{1}{\pi}\frac{\Gamma(\mu,k_z)}{\left(\mu - \epsilon_{n,l}\right)^2+\left[\Gamma(\mu,k_z)\right]^2}.
\end{equation}
In view of the Fermi arc dissipation due to transitions from the surface into the bulk~\cite{Gorbar-Sukhachov:2016}, we introduced a quasiparticle width for the surface states in addition to the bulk ones. The quasiparticle width of the Fermi arc states increases as the electric chemical potential and frequency grow because the phase space for the surface-bulk transitions rises as the area of the Fermi surface increases. By following Ref.~\cite{Gorbar-Sukhachov:2016}, we assume that $\Gamma(\mu,k_z)\approx\Gamma(\mu)=\Gamma_0 |\mu|/(\gamma m)$, where $\Gamma_0$ is a numerical constant. Furthermore, even without transitions to the bulk states (which is the case for small $\mu$ and $R$), the dissipative nature of the Fermi arcs can be explained by the fact that the finite size effects allow for a small but finite dispersion of their energy levels. This also signifies that the conventional point of view stating that the Fermi arcs are 1D nondissipative states should be considered carefully in real finite-size materials.

The integral over the azimuthal angle $\varphi^{\prime}$ can be straightforwardly performed. In particular,
it is easy to show that the terms $\propto M_{12}M_{21}^{\prime}$ and $\propto M_{21}M_{12}^{\prime}$, i.e., the third term in the curly brackets in Eq.~(\ref{Kubo-sigma-zz-DC}), vanish after the integration over $\varphi^{\prime}$. As expected, the resulting longitudinal conductivity becomes independent of $\varphi$. The integral over $r^{\prime}$ can be performed by using the corresponding formulas in Appendix~\ref{sec:app-Kubo-formulas}. Since the resulting expressions are cumbersome, we do not present them here.

Finally, let us briefly discuss the conductivity in Dirac semimetal nanowires. In such a case one should necessarily take into account the contribution from the lower block. Since the wave functions for the lower block are complex conjugated wave functions of the upper block, the corresponding contribution to the DC conductivity is the same as for the upper block. Therefore, the DC conductivity for Dirac semimetal nanowires is given by expression (\ref{Kubo-sigma-zz-DC}) with the additional factor $2$. It is worth noting that if there are terms that break the $\mathcal{P}$ and $\mathcal{T}$ symmetries, e.g., $\tilde{\mu}\sigma_z\otimes\mathds{1}_2$ and $\gamma m_1 \sigma_z\otimes\sigma_z$, then the effective electric chemical potentials and Weyl node separations in the upper and lower blocks become different. This will result in a nonvanishing magnetization current density $j_{\varphi}$ as well as nonequal contributions to the conductivity of the upper and lower blocks.

\subsection{Results}
\label{sec:Kubo-results}

In this subsection, we discuss the results for the DC conductivity. We concentrate primarily on the contribution from the upper block, which describes a $\mathcal{T}$ symmetry broken Weyl semimetal with two Weyl nodes. The conductivity of a Dirac semimetal nanowire in the model at hand will be simply doubled (see the corresponding discussion at the end of the previous subsection).

\begin{figure*}[!ht]
\begin{center}
\includegraphics[width=0.32\textwidth]{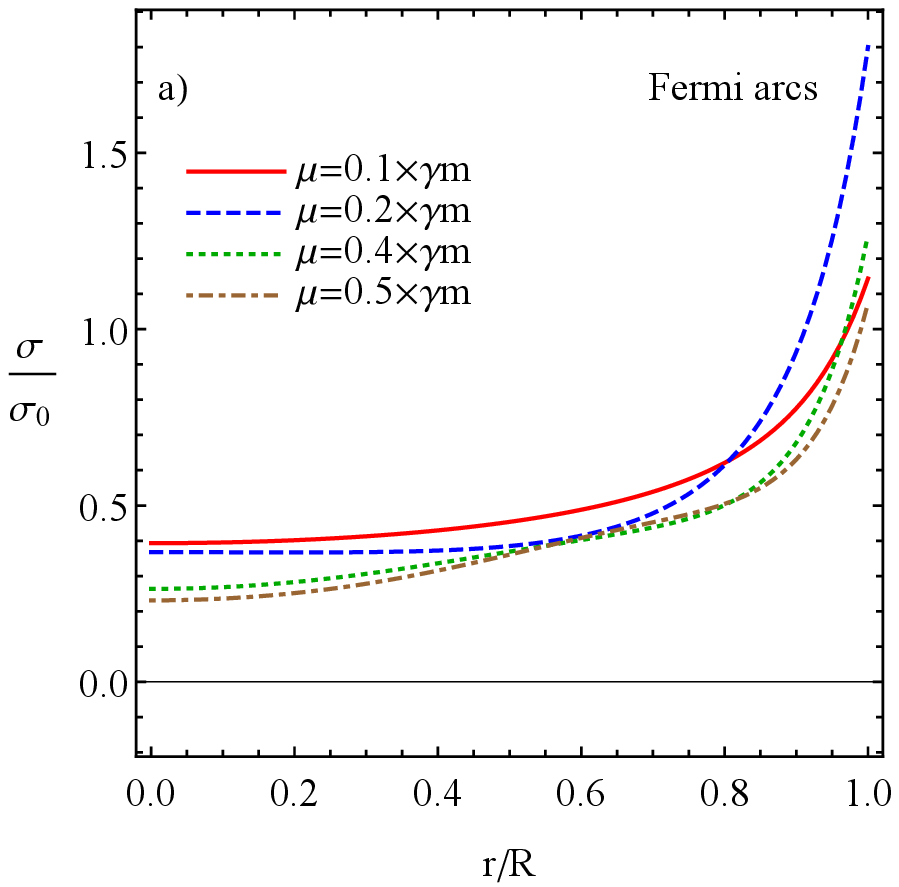}\hfill
\includegraphics[width=0.32\textwidth]{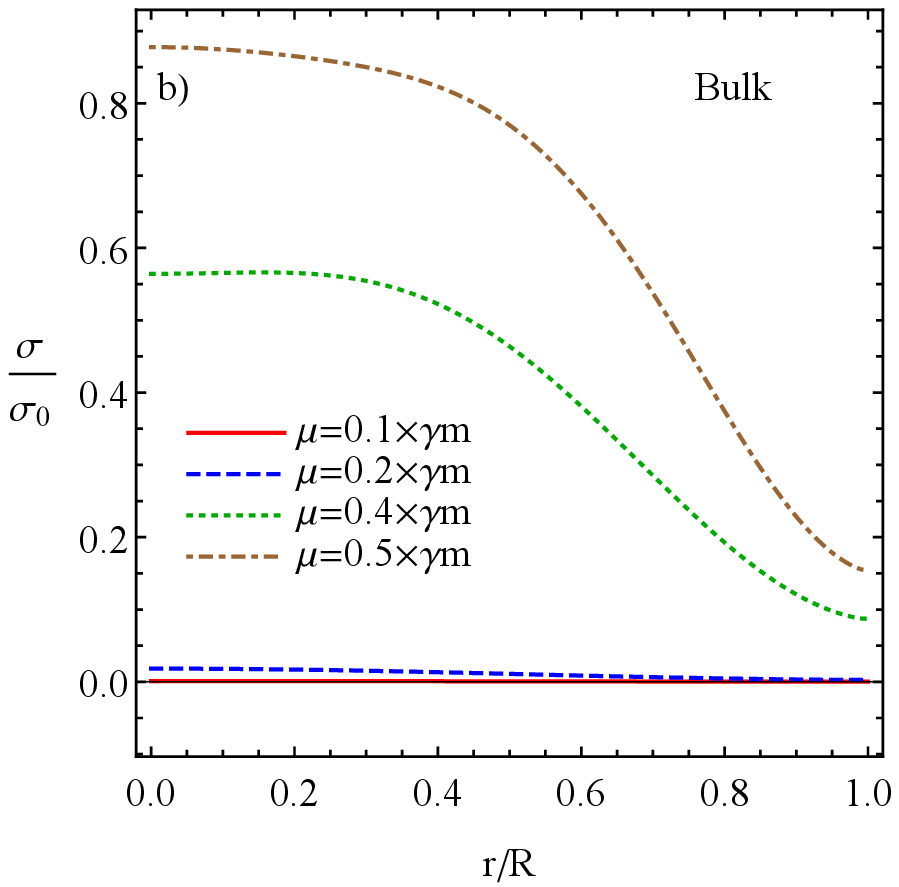}\hfill
\includegraphics[width=0.32\textwidth]{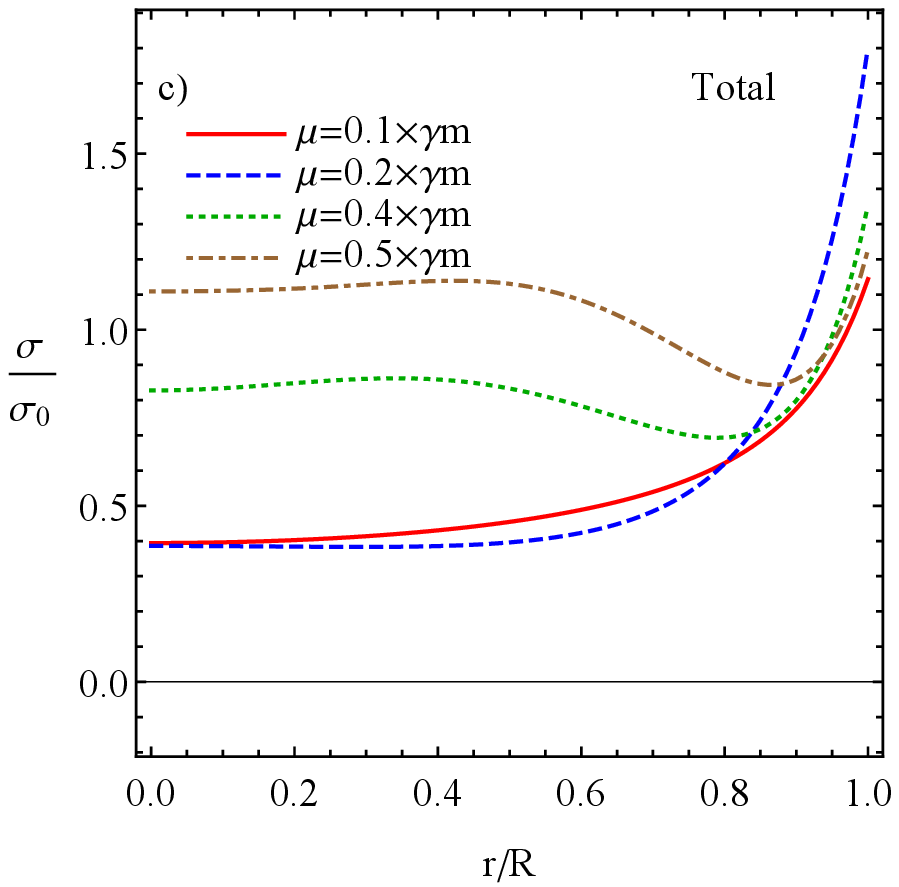}
\end{center}
\caption{
The longitudinal DC conductivity $\sigma$ integrated over the azimuthal angle $\varphi$
as a function of $r$ for several values of the electric chemical potential $\mu$. Panels (a) and (b) describe the contributions of the Fermi arc and bulk states, respectively. The total conductivity is presented in panel (c).
The conductivity is normalized on $\sigma_0=e^2\gamma^2m^{3/2}/v^2$ in all panels. Only the upper block (\ref{model-H-Dirac-up}) of the $4\times4$ Dirac Hamiltonian is considered, therefore, the results are valid for a $\mathcal{T}$ symmetry broken Weyl semimetal. In addition, we set $R=10\,v/(\gamma m)$, $\Gamma_0=0.2$, $T\to0$, and $C_0=C_1=0$.
}
\label{fig:Kubo-sigma-2x2-up-DC-separate}
\end{figure*}

To start with, let us consider the separate contributions from the surface and bulk states to the DC conductivity. The corresponding spatial profiles are shown in Figs.~\ref{fig:Kubo-sigma-2x2-up-DC-separate}(a) and \ref{fig:Kubo-sigma-2x2-up-DC-separate}(b), respectively, for several values of $\mu$.
Note that we integrated over $\varphi$ in both panels that gives the additional factor $2\pi$.
As expected, the contribution to the electric conductivity from the Fermi arcs significantly rises at the surface and has a nonmonotonic dependence on the electric chemical potential $\mu$. On the other hand, the bulk states allow for the electric conductivity that is large near the center of the wire with a sharp decrease at the surface. The corresponding contribution to $\sigma$ always rises with $\mu$.
At small $\mu$, i.e., $\mu=0.1\,\gamma m$ and $\mu=0.3\,\gamma m$, only the Fermi arcs are filled among the states with positive energy. Therefore, the rise of the total conductivity presented in Fig.~\ref{fig:Kubo-sigma-2x2-up-DC-separate}(c) near the surface is very pronounced. For larger $\mu$, i.e., $\mu=0.4\,\gamma m$ and $\mu=0.5\,\gamma m$, the bulk states also provide noticeable contribution to the transport. Moreover, since the phase space for the electron scattering increases with $\mu$, the relative contribution of the surface conductivity determined mainly by the Fermi arcs becomes suppressed compared to the contribution due to the bulk states. It is worth noting, however, that the quantitative analysis of the relative contributions from the surface and bulk localized states requires a more precise treatment of disorder.

\begin{figure*}[!ht]
\begin{center}
\includegraphics[width=0.45\textwidth]{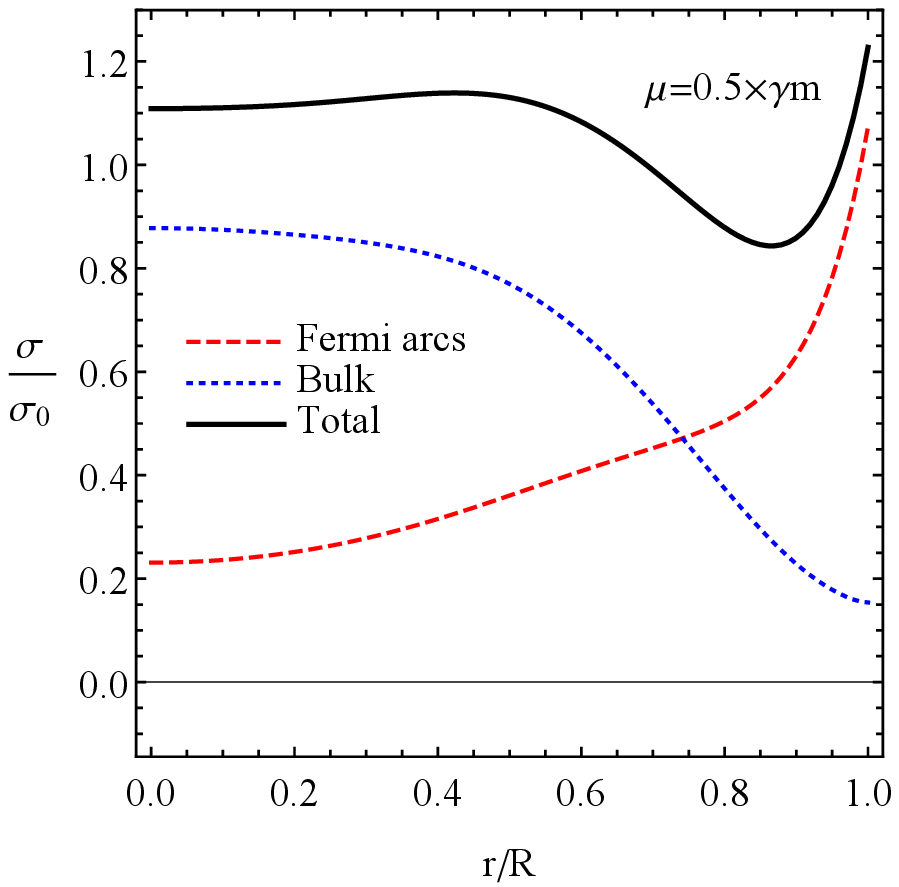}\hfill
\includegraphics[width=0.45\textwidth]{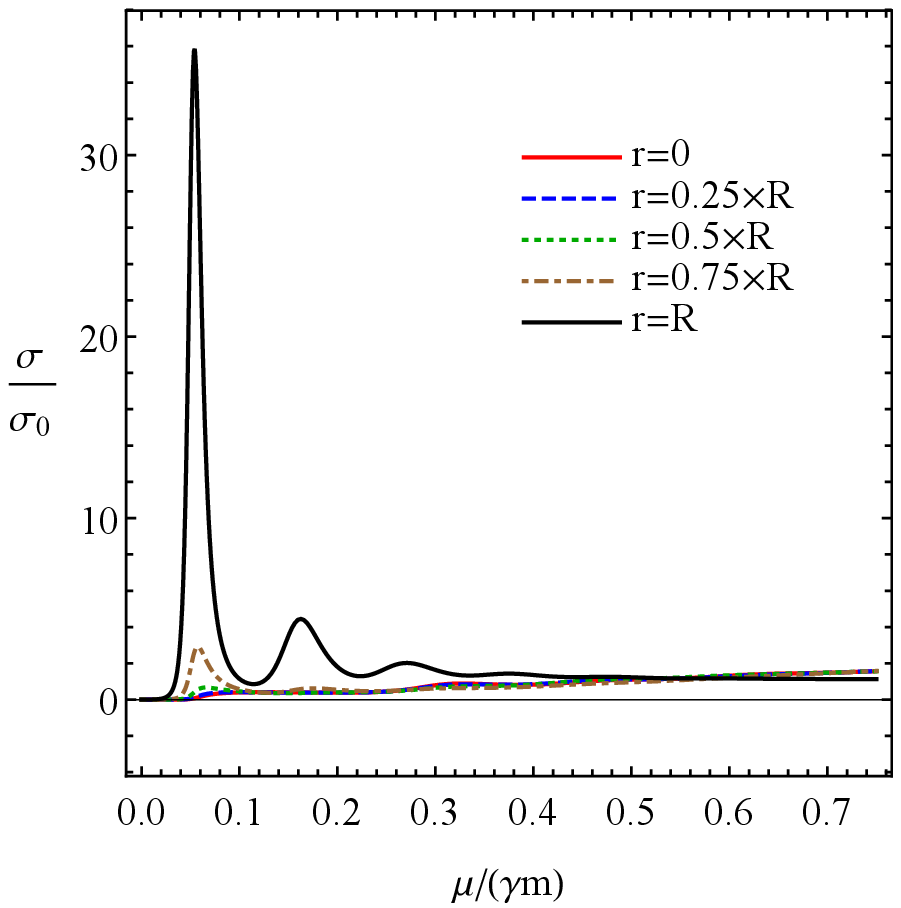}
\end{center}
\caption{
Left panel: The contributions to the longitudinal DC conductivity integrated over the azimuthal angle $\varphi$ from the surface (red dashed line) and bulk (blue dotted line) states as a function of $r$ for $\mu=0.5\,\gamma m$. Black solid line corresponds to the total conductivity.
Right panel: The total longitudinal DC conductivity $\sigma$ integrated over the azimuthal angle $\varphi$
aa a function of $\mu$ for several values of the coordinate $r$. The conductivity is normalized on $\sigma_0=e^2\gamma^2m^{3/2}/v^2$ in both panels. Only the upper block (\ref{model-H-Dirac-up}) of the $4\times4$ Dirac Hamiltonian is considered, therefore, the results are valid for a $\mathcal{T}$ symmetry broken Weyl semimetal. In both panels, we set $R=10\,v/(\gamma m)$, $\Gamma_0=0.2$, $T\to0$, and $C_0=C_1=0$.
}
\label{fig:Kubo-sigma-2x2-up-DC}
\end{figure*}

In order to understand better a nontrivial interplay of the surface and bulk states in the transport properties of the Dirac and Weyl semimetals nanowires, we compare the corresponding contributions for $\mu=0.5\,\gamma m$ in the left panel of Fig.~\ref{fig:Kubo-sigma-2x2-up-DC}. It is noticeable that the qualitatively different profiles of the surface and bulk conductivity contributions lead to the nonmonotonic dependence of $\sigma$ on $r$ when the electric chemical potential is sufficiently high and both surface and bulk states are populated. Indeed, the sharp increase of the total conductivity at the surface is followed by a well-pronounced minimum. Deeper in the bulk, the conductivity rises again. If observed, such a nontrivial profile might be a definite transport signature of the Fermi arcs surface states in nanowires of Weyl and certain Dirac semimetals.

The dependence of the DC conductivity on the electric chemical potential $\mu$ at a few values of the radial coordinate $r$ is shown in the right panel of Fig.~\ref{fig:Kubo-sigma-2x2-up-DC}. One can clearly see that the longitudinal conductivity has a few peaks as a function of $\mu$. We traced back the origin of these peaks to the Fermi arcs. Indeed, as follows from the right panel of Fig.~\ref{fig:spectrum-2x2-large-R}, the position of the peaks coincide with the energy of the Fermi arc states. Furthermore, the height of the peaks significantly increases at the surface, which also supports their relation to the Fermi arcs.
In addition, it is noticeable that the surface conductivity generically tends to decrease with the electric chemical potential. This is related to the fact that the surface-bulk transition rate grows with $\mu$ leading to higher Fermi arc dissipation. On the other hand, the bulk conductivity rises with the Fermi level despite the increase of the quasiparticle width.

Finally, we suggest a few ways to experimentally verify the proposed effects. Clearly, the most direct means to investigate the nontrivial spatial profile of the conductivity is to measure the current density distribution at the ends of the wire. In theory, this can be performed by attaching several small contacts and measuring the corresponding currents. Such an approach, however, might be difficult to realize in practice for thin wires.
Further, the nontrivial contribution of the Fermi arcs in nanowires can be probed via the dependence on the electric chemical potential (see the right panel in Fig.~\ref{fig:Kubo-sigma-2x2-up-DC}).
Unlike the bulk-dominated regime, where the conductivity gradually increases with $\mu$, the Fermi arc contributions allow for peaks in the conductivity. The corresponding dependence could be in principle studied by doping the wire and/or investigating different samples with the same radius.
Another way to probe the interplay of the surface and bulk states in the DC transport is to investigate the scaling of the conductance $G$ in nanowires with their radius. The conductance is defined as
\begin{eqnarray}
\label{Kubo-results-Conductance}
G =  \int_0^{2\pi}d\varphi \int_0^{R}r\,dr\, \sigma(r, \varphi) = 2\pi \int_0^{R}r\,dr\, \sigma(r).
\end{eqnarray}

\begin{figure*}[!ht]
\begin{center}
\includegraphics[width=0.45\textwidth]{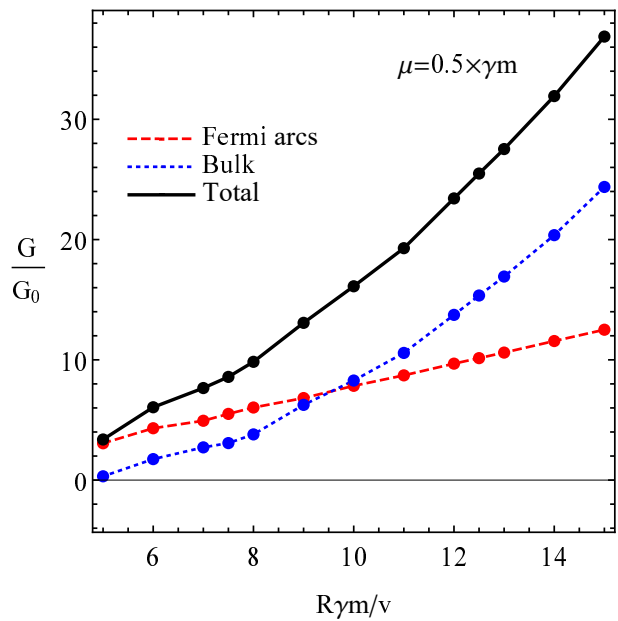}
\end{center}
\caption{
The contributions to the longitudinal DC conductance $G$ defined in Eq.~(\ref{Kubo-results-Conductance}) as functions of the wire radius $R$ for $\mu=0.5\,\gamma m$. Red dashed and blue dotted lines correspond to the Fermi arc and bulk contributions, respectively. The total conductance is denoted by a solid black line. Only the upper block (\ref{model-H-Dirac-up}) of the $4\times4$ Dirac Hamiltonian is considered, therefore, the results are valid for a $\mathcal{T}$ symmetry broken Weyl semimetal. In addition, we set $G_0=\pi e^2/\sqrt{m}$, $\Gamma_0=0.2$, $T\to0$, and $C_0=C_1=0$.
}
\label{fig:Kubo-Conductance-2x2-up-DC-all}
\end{figure*}

The contribution to the DC conductance from the surface and bulks states as well as the total conductance are shown in Fig.~\ref{fig:Kubo-Conductance-2x2-up-DC-all}. As one can see, the Fermi arc contribution (red dashed line) to the conductance scales approximately as a perimeter of the wire, i.e., linearly with radius $G\sim R$. Such a behavior is indeed expected for the surface-localized contribution where $G\sim R \sigma(R)$. On the other hand, the bulk contribution shown by the blue dotted line demonstrates a quadratic dependence on the radius $R$. This is indeed expected for bulk conductors with almost unform conductivity where conductance is determined by the cross-section $G\sim R^2 \sigma(0)$.
Therefore, we suggest to use a relatively thick wire of Weyl or Dirac semimetal with $R\sim100~\mbox{nm}$ and change its radius via, e.g., the focused ion beam method. It should be possible, also, to use natural crystals of different radius. For example, the Dirac semimetal Cd$_3$As$_2$ could be routinely crystalized in a wire-like form~\cite{Li-Yu-Cd3As2:2015,Wang-Liao:2016,Bayogan-Jung:2019}.
In this case, however, one should be careful because the position of the Fermi level might depend on the sample too.
Either way, similarly to the experimental studies performed in Ref.~\cite{Gooth-Gotsmann:2018}, it should be possible to investigate the scaling of the conductance with the radius $R$.

\section{Summary}
\label{sec:Summary}

In this study, we investigated the DC transport properties of nanowires of Dirac and Weyl semimetals paying special attention to the interplay of
the surface and bulk modes. In order to distinguish the corresponding contributions, we determined the wave functions and calculated the
profiles of the charge and current densities. In addition, by using the Kubo linear response theory, the local DC conductivity was analyzed.

For cylindrical nanowires where the vector $\mathbf{b}$, which defines the Weyl nodes separation direction, is parallel to the axis of the wire, the wave functions and the characteristic equation for the energy spectrum were found analytically and agree with those in the literature. As expected, the energy spectrum of the Fermi arc states becomes discrete due to the effects of spatial confinement. In contrast to a semi-infinite slab with flat Fermi arc bands, the energy levels obtain a weak quadratic dispersion with respect to the momentum along the cylinder axis.
This dispersion becomes particularly noticeable 
near the bulk nodes. Therefore, strictly speaking, the Fermi arcs become dissipative in real finite-size materials even if the electron scattering to the bulk is ignored.

We found that the electric charge distribution in nanowires is nonuniform. Due to the Fermi arc states, a significant amount of charge is located at the surface. The contribution of the bulk states increases with the growth of the electric chemical potential and becomes dominant for a sufficiently large value of the latter.
As to the magnetization current in Weyl semimetal nanowires, we found that it does not vanish in an equilibrium state when the external fields are absent. Only the azimuthal component of this current is nontrivial, however. Like the electric charge density, the magnetization current is primarily connected with the surface states and monotonically decreases into the bulk. On the other hand, the bulk contribution oscillates and diminishes away from the center of nanowires. The direction of the azimuthal component of the magnetization current is determined by the chiral shift. Indeed, we checked explicitly that this current runs in the opposite directions for the two copies of Weyl semimetal that constitute the Dirac semimetal at hand. Therefore, since the time-reversal symmetry is preserved in Dirac semimetals, the total magnetization current vanishes.

By using the Kubo linear response theory, we calculated the longitudinal conductivity for the DC response. The DC conductivity $\sigma$ has a nontrivial profile, which strongly depends on the electric chemical potential. By explicitly separating the surface and bulk contributions, we showed that if the Fermi level intersects only energies of the surface states, then the conductivity is high at the surface and monotonously diminishes in the bulk. The contribution of the bulk states grows with the Fermi energy producing a noticeable increase of the conductivity inside nanowires and decreasing the relative surface contribution. In part, this is related to the increase of the surface-bulk scattering. For sufficiently large electric chemical potential, the spatial profile of $\sigma$ is nonmonotonic where a sharp rise of the conductivity at the surface is followed by a local minimum. Due to the contribution of the bulk states, the conductivity rises again deep in the bulk.
We found also that the surface conductivity shows well-pronounced peaks when the electric chemical potential intersects the Fermi arc energy levels. The surface origin of these peaks also follows from the fact that their amplitude significantly grows from the center of nanowires to their surface.
If experimentally probed, the nontrivial distribution of the electric conductivity will be a definite signature of the interplay of the surface and the bulk states in Weyl and Dirac semimetals nanowires. In addition, the bulk- and surface-dominated regimes could be distinguished via the conductance measurements. In the former case, the conductance scales as the square of the wire radius $R$. However, it changes to the linear dependence when the Fermi arcs dominates the transport.

Finally, let us discuss the key limitations of this study. We treated disorder phenomenologically by introducing a finite quasiparticle width for both surface and bulk states. The rigorous investigation of the disorder scattering in nanowires is a very interesting problem that deserves an in-depth investigation, which is, however, beyond the scope of this study.
In general, the irregularities of the wire shape, which are present in real samples, should be also taken into account. We believe, however, that they will not change the main qualitative conclusions of this study.
In addition, we leave the investigation of the effects of various $\mathcal{T}$ and $\mathcal{P}$ symmetry breaking terms on the nanowire transport for future research.

\begin{acknowledgments}
We are grateful to I.~A.~Shovkovy, J.~H.~Bardarson, and V.~Kaladzhyan for useful discussions.
P.O.S. was supported by the VILLUM FONDEN via the Centre of Excellence for Dirac Materials (Grant No.~11744), the European Research Council under the European Unions Seventh Framework Program Synergy HERO, and the Knut and Alice Wallenberg Foundation KAW 2018.0104. M.V.R. acknowledges funding by the Deutsche Forschungsgemeinschaft (DFG) under Germany's Excellence Strategy - EXC-2123/1. The work of E.V.G. was supported partially by the Ukrainian State Foundation for Fundamental Research.
\end{acknowledgments}

\appendix

\section{Wave function in the wire}
\label{sec:app-wave-functions}

In this section, we determine the wave functions in Dirac and Weyl semimetal nanowires. It suffices to consider only the upper
block given by Eq.~(\ref{model-H-Dirac-up}) because the wave functions for the lower block are simply complex conjugated ones
$\psi^{-}=(\psi^{+})^{*}$. Seeking the wave functions inside the wire ($r<R$) in the form (\ref{model-cylinder-psi-Dirac-up-r<R}), i.e.,
\begin{equation}
\label{app-wf-psi-Dirac-up-r<R}
\psi^{+} = \left(
                \begin{array}{c}
                  \rho_-(r) e^{in\varphi} \\
                 \rho_+(r) e^{i(n-1)\varphi} \\
                \end{array}
              \right),
\end{equation}
and squaring the eigenvalue equation $H_{2\times2}^{+}\psi = \epsilon \psi$, we obtain
\begin{equation}
\label{app-wf-H2}
(H_{2\times2}^{+})^2 = \left(
                        \begin{array}{cc}
                          \gamma^2(k_z^2-m)^2-v^2\left(\frac{\partial^2}{\partial r^2}+\frac{1}{r}\frac{\partial}{\partial r}+\frac{1}{r^2}\frac{\partial^2}{\partial \varphi^2}\right) & 0 \\
                          0 & \gamma^2(k_z^2-m)^2-v^2\left(\frac{\partial^2}{\partial r^2}+\frac{1}{r}\frac{\partial}{\partial r}+\frac{1}{r^2}\frac{\partial^2}{\partial \varphi^2}\right) \\
                        \end{array}
                      \right).
\end{equation}
This leads to the equation
\begin{equation}
\left[\gamma^2(k_z^2-m)^2-v^2\left(\frac{\partial^2}{\partial r^2}+\frac{1}{r}\frac{\partial}{\partial r}+\frac{1}{r^2}\frac{\partial^2}{\partial \varphi^2}\right)\right]\rho_{\zeta}(r) e^{i\left[n-\theta(\zeta)\right]\varphi}=\tilde{\epsilon}^2 \rho_{\zeta}(r) e^{i\left[n-\theta(\zeta)\right]\varphi},
\label{squared-Hamiltonian-equation}
\end{equation}
where $\zeta=\pm$, $\theta(x)$ is the Heaviside function, and $\tilde{\epsilon} = \epsilon - C_0 -C_1k_z^2$.
Equation (\ref{squared-Hamiltonian-equation}) implies
\begin{equation}
\label{app-wf-rho-equation}
\frac{d^2 \rho_{\zeta}(r)}{d r^2}+\frac{1}{r}\frac{d \rho_{\zeta}(r)}{d r}+\left\{\frac{s_{\epsilon}}{r_0^2}-\frac{\left[n-\theta(\zeta)\right]^2}{r^2}\right\}\rho_{\zeta}(r)=0,
\end{equation}
where
\begin{equation}
\label{app-wf-r0-def}
r_0 = \frac{v}{\sqrt{\left|\tilde{\epsilon}^{2}-\gamma^2(k_z^2-m)^2\right|}}, \quad\quad
s_{\epsilon} = \mbox{sgn}{\left[\tilde{\epsilon}^{2}-\gamma^2(k_z^2-m)^2\right]}.
\end{equation}

Equation~(\ref{app-wf-rho-equation}) has a normalizable solution either in terms of the Bessel $J_n(x)$ (for $s_{\epsilon}>0$) or modified Bessel $I_n(x)$ (for $s_{\epsilon}<0$) functions of the first kind. In particular, we have
\begin{eqnarray}
\label{app-wf-rho-s-plus}
s_{\epsilon}>0: &\quad& \rho_{\zeta}(r) = a_{\zeta} \, J_{n-\theta(\zeta)}\left(\frac{r}{r_0}\right),\\ 
\label{app-wf-rho-s-minus}
s_{\epsilon}<0: &\quad& \rho_{\zeta}(r) = b_{\zeta} \, I_{n-\theta(\zeta)}\left(\frac{r}{r_0}\right).
\end{eqnarray}

The relation between the coefficients $a_{+}$ and $a_{-}$ can be found by substituting $\rho_{\zeta}(r)$ given in Eq.~(\ref{app-wf-rho-s-plus}) into the eigenvalue equation $H_{2\times2}^{+}\psi=\epsilon \psi$. For $s_{\epsilon}>0$, we find
\begin{equation}
\label{app-wf-H2x2-plus-eigensystem}
\left(
  \begin{array}{c}
    e^{in\varphi} \left[\gamma\left(k_z^2-m\right) \rho_-(r) - iv\left(\frac{d}{dr}-\frac{n-1}{r}\right)\rho_+(r)\right] \\
    e^{i(n-1)\varphi} \left[-iv\left(\frac{\partial}{\partial r}+\frac{n}{r}\right)\rho_-(r)-\gamma\left(k_z^2-m\right) \rho_+(r)\right] \\
  \end{array}
\right) = \tilde{\epsilon} \left(
                             \begin{array}{c}
                               e^{in\varphi} \rho_-(r) \\
                               e^{i(n-1)\varphi} \rho_+(r) \\
                             \end{array}
                           \right).
\end{equation}
Then, by using the recurrence relation for the Bessel functions $\left(d/dx+n/x\right)J_n(x)=J_{n-1}(x)$, we obtain
\begin{equation}
\label{app-wf-a-plus-a-minus}
a_+=-ia_- \, F(\tilde{\epsilon}), 
\end{equation}
where
\begin{equation}
\label{app-wf-F-def}
F(\tilde{\epsilon}) = \frac{\sqrt{\left|\tilde{\epsilon}^{2}-\gamma^2(k_z^2-m)^2\right|}}{\tilde{\epsilon}+\gamma(k_z^2-m)}.
\end{equation}

The normalization condition of wave functions reads as
\begin{eqnarray}
\label{app-wf-normalization}
\delta_{n,n^{\prime}} &=& \int_{0}^{R}rdr \int_{0}^{2\pi}d\varphi\, \psi_{n}^{\dag}(r)\psi_{n^{\prime}}(r) =|a_-|^2 \, \int_0^{2\pi} e^{i(n-n^{\prime})\varphi} d\varphi \, \int_0^R rdr\, \nonumber\\
&\times&\left[J_{n^{\prime}}\left(\frac{r}{r_0}\right)J_n\left(\frac{r}{r_0}\right) +
\left|F(\tilde{\epsilon})\right|^2
J_{n^{\prime}-1}\left(\frac{r}{r_0}\right)J_{n-1}\left(\frac{r}{r_0}\right)\right].
\end{eqnarray}
The integral over the angle gives the Kronecker symbol $\delta_{n,n^{\prime}}$. By using formula 1.8.3.12 in Ref.~\cite{Prudnikov-Marichev:book} and the recurrence relation for the derivative of $J_n(x)$, we obtain
\begin{equation}
\label{app-wf-Jna-def}
\mathcal{J}(n,r_0)=\int_0^Rrdr \, J_n^2\left(\frac{r}{r_0}\right)=\frac{R^2}{2}\left[J_n^2\left(\frac{R}{r_0}\right)+J_{n+1}^2\left(\frac{R}{r_0}\right)\right] -nRr_0J_n\left(\frac{R}{r_0}\right)J_{n+1}\left(\frac{R}{r_0}\right).
\end{equation}
As a result, the normalization constant for wave functions equals
\begin{equation}
\label{app-wf-A-plus-def}
A_{+}\equiv |a_{-}|=\frac{1}{\sqrt{2\pi\left[\mathcal{J}(n,r_0) + F^2(\tilde{\epsilon})
\mathcal{J}(n-1,r_0)\right]}}
\end{equation}
Note that, in principle, one should integrate over the whole space, i.e., $r\geq0$ in Eq.~(\ref{app-wf-normalization}). However, as can be
straightforwardly shown by using the results in Appendix~\ref{sec:app-BC}, the contribution of the vacuum part of the wave
function is $\propto 1/\tilde{m}$ and is negligible for $\tilde{m}\to\infty$.

The wave functions and their normalization for $s_{\epsilon}<0$ can be determined in a similar way. In such a case,
\begin{equation}
\label{app-wf-b-plus-b-minus}
b_+=-i b_- \, F(\tilde{\epsilon}).
\end{equation}
Then, by using formula 1.11.3.4 in Ref.~\cite{Prudnikov-Marichev:book} and a recurrence relation for the derivative of $I_n(x)$, we derive
\begin{equation}
\label{app-wf-Ina-def}
\mathcal{I}(n,r_0)=\int_0^Rrdr\,I_n^2\left(\frac{r}{r_0}\right) =\frac{R^2}{2}\left[I_n^2\left(\frac{R}{r_0}\right)-I_{n+1}^2\left(\frac{R}{r_0}\right)\right] -nRr_0I_n\left(\frac{R}{r_0}\right)I_{n+1}\left(\frac{R}{r_0}\right).
\end{equation}
Therefore, the normalization coefficient is given by
\begin{equation}
\label{wave-functions-A-fin-2}
A_{-}\equiv |b_{-}|=\frac{1}{\sqrt{2\pi\left[\mathcal{I}(n,r_0) +F^2(\tilde{\epsilon})
\mathcal{I}(n-1,r_0)\right]}}.
\end{equation}
Note that the corresponding results obtained in this section agree with those derived in Ref.~\cite{Kaladzhyan-Bardarson:2019}.

\section{Derivation of the boundary conditions}
\label{sec:app-BC}

In this section, we derive the boundary condition given in Eq.~(\ref{model-cylinder-BC}) in the main text. As in the previous
Section, it is sufficient to consider only the upper block in Hamiltonian~(\ref{model-H-Dirac}) in the main text. The boundary condition for
the lower block is the same.

In vacuum, i.e., for $r>R$, one should replace $m\to-\tilde{m}$. Therefore, we can use an ansatz that is similar to that in Eq.~(\ref{app-wf-psi-Dirac-up-r<R}) for wave functions outside the wire
\begin{equation}
\label{app-model-cylinder-psi-Dirac-up-r>R}
\tilde{\psi}^{+} = \left(
                \begin{array}{c}
                  \tilde{\rho}_-(r) \, e^{in\varphi} \\
                  \tilde{\rho}_+(r) \, e^{i(n-1)\varphi} \\
                \end{array}
              \right) e^{-\kappa (r-R)}.
\end{equation}
Then solving the eigenvalue equation $\tilde{H}^{+}_{2\times2}\tilde{\psi}^{+}=\epsilon\tilde{\psi}^{+}$, where
$\tilde{H}^+_{2\times 2}$ is the same as $H^+_{2\times 2}$ but with the replacement $m\to-\tilde{m}$, we obtain
\begin{eqnarray}
\label{app-model-cylinder-Hpsi-1}
&&e^{-\kappa (r-R)}  e^{in\varphi} \left[\gamma(k_z^2+\tilde m) \tilde{\rho}_-(r) - iv\left(\frac{\partial}{\partial r}-\frac{n-1}{r}-\kappa\right)\tilde{\rho}_+(r)\right] = \tilde{\epsilon} e^{-\kappa (r-R)} e^{in\varphi} \tilde{\rho}_-(r),\\
\label{app-model-cylinder-Hpsi-2}
&&e^{-\kappa (r-R)}  e^{i(n-1)\varphi} \left[-iv\left(\frac{\partial}{\partial r}+\frac{n}{r}-\kappa\right)\tilde{\rho}_-(r)-\gamma(k_z^2+\tilde m) \tilde{\rho}_+(r)\right] = \tilde{\epsilon} e^{-\kappa (r-R)} e^{i(n-1)\varphi} \tilde{\rho}_+(r).
\end{eqnarray}
Since the behavior of wave functions far from the wire is determined mainly by the dominant exponential factor, we can ignore the
spatial dependence of $\tilde{\rho}_{+}(r)$ and $\tilde{\rho}_{-}(r)$ as well as neglect terms $\propto 1/r$. The nontrivial solutions of the
resulting system
\begin{eqnarray}
\label{app-model-cylinder-Hpsi-1-kappa}
&&e^{-\kappa (r-R)}  e^{in\varphi} \left[\gamma(k_z^2+\tilde m) \tilde{\rho}_- + iv\kappa\tilde{\rho}_+\right] = \tilde{\epsilon} e^{-\kappa (r-R)} e^{in\varphi} \tilde{\rho}_-,\\
\label{app-model-cylinder-Hpsi-2-kappa}
&&e^{-\kappa (r-R)}  e^{i(n-1)\varphi} \left[iv\kappa\tilde{\rho}_- -\gamma(k_z^2+\tilde m) \tilde{\rho}_+(r)\right] = \tilde{\epsilon} e^{-\kappa (r-R)} e^{i(n-1)\varphi} \tilde{\rho}_+
\end{eqnarray}
exist for
\begin{equation}
\label{model-cylinder-kappa-def}
\kappa =\frac{\sqrt{\gamma^2(k_z^2+\tilde{m})^2-\tilde{\epsilon}^2}}{v} \approx \frac{\gamma \tilde{m}}{v}.
\end{equation}
By substituting the above expression for $\kappa$ into the system (\ref{app-model-cylinder-Hpsi-1-kappa}) and (\ref{app-model-cylinder-Hpsi-2-kappa}), the following relation can be derived:
\begin{equation}
\label{app-model-cylinder-BC}
\tilde{\rho}_-(r)+i\tilde{\rho}_+(r)=0.
\end{equation}
The final step is to match the vacuum wave function (\ref{app-model-cylinder-psi-Dirac-up-r>R}) with its bulk
counterpart (\ref{app-wf-psi-Dirac-up-r<R}) at the boundary $r=R$, which leads to $\tilde{\rho}_{-}(R)=\rho_{-}(R)$ and
$\tilde{\rho}_{+}(R)=\rho_{+}(R)$. Then the boundary condition given in Eq.~(\ref{model-cylinder-BC}) in the main text trivially follows from
Eq.~(\ref{app-model-cylinder-BC}). This boundary condition agrees with that obtained in Ref.~\cite{Kaladzhyan-Bardarson:2019}.
In addition, we note that a general form of the boundary conditions for the wave functions and a dispersion relation in cylindrical wires of Weyl semimetals was derived in Ref.~\cite{Erementchouk-Mazumder:2018}. In particular, a generalized analog of Eq.~(\ref{app-model-cylinder-BC}) depends on the angle between the quasiparticle velocity and the axis of the wire.

\section{Conductivity for wires of different radiuses}
\label{sec:app-sigma-few-R}

\begin{figure*}[!ht]
\begin{center}
\includegraphics[width=0.45\textwidth]{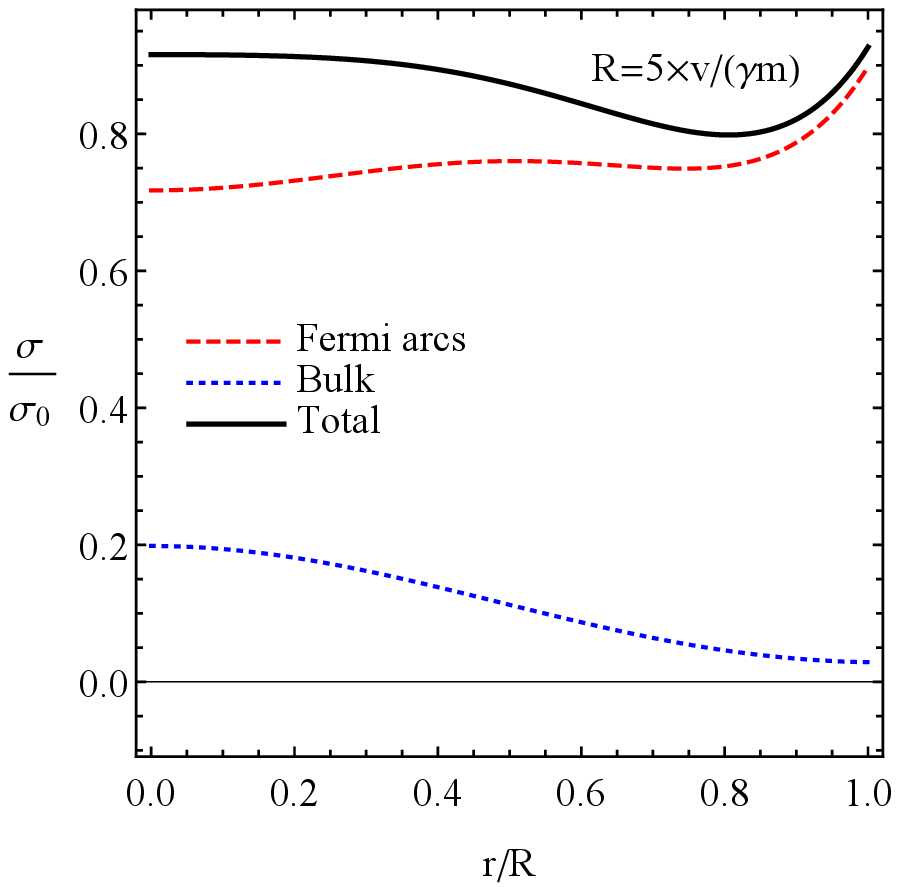}\hfill
\includegraphics[width=0.45\textwidth]{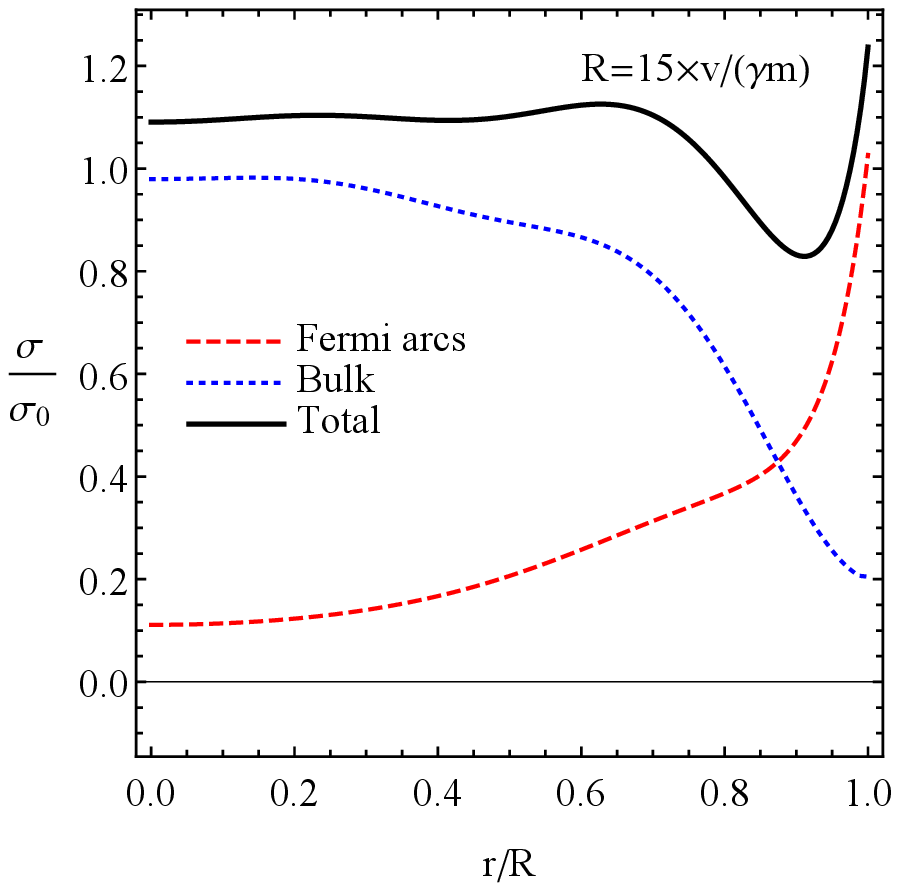}
\end{center}
\caption{The contributions to the longitudinal DC conductivity $\sigma$ integrated over the azimuthal angle $\varphi$ from the surface (red dashed lines) and bulk (blue dotted lines) states as a function of $r$ for $R=5\,v/(\gamma m)$ (left panel) and $R=15\,v/(\gamma m)$ (right panel). Black solid line corresponds to the total conductivity. The conductivity is normalized on $\sigma_0=e^2\gamma^2m^{3/2}/v^2$ in both panels. Only the upper block (\ref{model-H-Dirac-up}) of the $4\times4$ Dirac Hamiltonian is considered, therefore, the results are valid for a $\mathcal{T}$ symmetry broken Weyl semimetal. In both panels, we set $\mu=0.5\,\gamma m$, $\Gamma_0=0.2$, $T\to0$, and $C_0=C_1=0$.
}
\label{fig:Kubo-sigma-2x2-few-R}
\end{figure*}

To demonstrate the evolution of the transport properties of nanowires, it is convenient to present a few results for wires of different radius $R$. The conductivities for $R=5\,v/(\gamma m)$ and $R=15\,v/(\gamma m)$ are shown in the left and right panels of Fig.~\ref{fig:Kubo-sigma-2x2-few-R}, respectively. As one can see, the Fermi arc contribution is dominant for small nanowires (left panel). The contribution of the bulk states gradually rises with the increase of the wire thickness (see the right panel in Fig.~\ref{fig:Kubo-sigma-2x2-few-R}).

\section{Useful formulas and relations}
\label{sec:app-Kubo-formulas}

In this section, some formulas useful for the calculations of integrals over $r^{\prime}$ in the conductivity are presented.

The integrals over $r^{\prime}$ for terms $M_{11}M_{11}^{\prime}$ and $M_{22}M_{22}^{\prime}$ (the matrix elements $M_{ij}$ are
defined in Eq.~(\ref{Kubo-A-matrix}) in the main text)
can be calculated by using formulas 1.8.3.10 and 1.8.3.12 in Ref.~\cite{Prudnikov-Marichev:book}
\begin{eqnarray}
\label{Kubo-1.8.3.10}
\int_0^{R} r dr\, J_n\left(\frac{r}{r_0}\right) J_n\left(\frac{r}{r_0^{\prime}}\right) &=& \frac{R (r_0^{\prime})^2 r_0^2}{(r_0^{\prime})^2-r_0^2} \left[\frac{1}{r_0} J_{n+1}\left(\frac{R}{r_0}\right) J_n\left(\frac{R}{r_0^{\prime}}\right) -\frac{1}{r_0^{\prime}} J_{n}\left(\frac{R}{r_0}\right) J_{n+1}\left(\frac{R}{r_0^{\prime}}\right)\right],\\
\label{Kubo-1.8.3.12}
\int_0^{R} r dr\, J_n^2\left(\frac{r}{r_0}\right) &=& \frac{R^2}{2} \left[J_n^2\left(\frac{R}{r_0}\right) -J_{n-1}\left(\frac{R}{r_0}\right)J_{n+1}\left(\frac{R}{r_0}\right)\right]\nonumber\\
&=& \frac{R^2}{2}\left[J_n^2\left(\frac{R}{r_0}\right)+J_{n+1}^2\left(\frac{R}{r_0}\right)\right]-nRr_0J_n\left(\frac{R}{r_0}\right)J_{n+1}\left(\frac{R}{r_0}\right).
\end{eqnarray}
For the modified Bessel functions of the first kind, the corresponding integrals read as
\begin{eqnarray}
\label{Kubo-1.11.3.3}
\int_0^{R} r dr\, I_n\left(\frac{r}{r_0}\right) I_n\left(\frac{r}{r_0^{\prime}}\right) &=& \frac{R (r_0^{\prime})^2 r_0^2}{(r_0^{\prime})^2-r_0^2} \left[\frac{1}{r_0} I_{n+1}\left(\frac{R}{r_0}\right) I_n\left(\frac{r}{r_0^{\prime}}\right) -\frac{1}{r_0^{\prime}} I_{n}\left(\frac{R}{r_0}\right) I_{n+1}\left(\frac{R}{r_0^{\prime}}\right)\right],\\
\label{Kubo-1.11.3.4}
\int_0^{R} r dr\, I_n^2\left(\frac{r}{r_0}\right) &=& -\frac{R^2}{2} \left[I_n^{\prime}\left(\frac{R}{r_0}\right) +\frac{1}{2}\left(R^2 + n^2 r_0^2\right) I_n^2 \left(\frac{R}{r_0}\right)\right] \nonumber\\
&=& \frac{R^2}{2}\left[I_n^2\left(\frac{R}{r_0}\right)-I_{n+1}^2\left(\frac{R}{r_0}\right)\right] -nRr_0I_n\left(\frac{R}{r_0}\right)I_{n+1}\left(\frac{R}{r_0}\right),
\end{eqnarray}
where formulas 1.11.3.3 and 1.11.3.4 in Ref.~\cite{Prudnikov-Marichev:book} were used.
For $s_{\epsilon}s_{\epsilon^{\prime}}<0$, we have
\begin{eqnarray}
\label{Kubo-1.11.5.1}
\int_0^{R} r dr\, J_n\left(\frac{r}{r_0}\right) I_n\left(\frac{r}{r_0^{\prime}}\right) &=& \frac{R (r_0^{\prime})^2 r_0^2}{(r_0^{\prime})^2+r_0^2} \left[\frac{1}{r_0} J_{n+1}\left(\frac{r}{r_0}\right) I_n\left(\frac{r}{r_0^{\prime}}\right) +\frac{1}{r_0^{\prime}} J_{n}\left(\frac{r}{r_0}\right) I_{n+1}\left(\frac{r}{r_0^{\prime}}\right)\right].
\end{eqnarray}

\end{document}